\documentclass[english,aps,prl,preprint, groupedaddress, amsmath]{revtex4-2}
\usepackage{ulem}
\usepackage{fancyhdr,graphicx}
\usepackage{bm}
\usepackage{color}
\usepackage{amsfonts}
\usepackage{amssymb}
\usepackage{times}
\usepackage{amsmath}
\usepackage{graphicx}
\usepackage{float}
\usepackage{parskip} 
\usepackage{natbib}
\usepackage[utf8]{inputenc}
\usepackage{braket}
\usepackage{ulem}
\usepackage{chemformula} 
\usepackage[colorlinks=true, citecolor=black]{hyperref}
\setcitestyle{journalcolor= blue}
\usepackage{chemformula} 
\usepackage[T1]{fontenc} 

\usepackage{amsmath} 

\usepackage{babel}

\newcounter{suppfigure}

\newenvironment{suppfigure}{
	\refstepcounter{suppfigure} 
	\renewcommand{\thefigure}{\thesuppfigure} 
	\begin{figure}
	}{
	\end{figure}
	\renewcommand{\thefigure}{\arabic{figure}} 
}

\begin{document}

\title{Controlling particle-hole symmetry of fractional quantum hall states in trilayer graphene}

\author{Simrandeep Kaur$^1$, Harsimran Singh$^1$, Kenji Watanabe$^3$, Takashi Taniguchi$^4$, Unmesh Ghorai$^5$, Manish Jain$^1$, Rajdeep Sensarma$^5$}
\author{Aveek Bid$^1$}
\email{aveek@iisc.ac.in}
\affiliation{$^1$Department of Physics, Indian Institute of Science, Bangalore 560012, India \\
	$^2$ Research Center for Functional Materials, National Institute for Materials Science, 1-1 Namiki, Tsukuba 305-0044, Japan \\
	$^4$ International Center for Materials Nanoarchitectonics, National Institute for Materials Science, 1-1 Namiki, Tsukuba 305-0044, Japan\\
	$^5$ Department of Theoretical Physics, Tata Institute of Fundamental Research, Homi Bhabha Road, Mumbai, 400005, India}

\begin{abstract}
	We present a detailed experimental study of the particle-hole symmetry (PHS)  of the fractional quantum Hall (FQH) states about half filling in a multiband system. Specifically, we focus on the lowest Landau level of the monolayer-like band of Bernal stacked trilayer graphene (TLG). In pristine TLG, the excitation energy gaps, Land\'e g-factor, effective mass, and disorder broadening of the odd-denominator FQH states are identical to their hole-conjugate counterpart. This precise PH symmetry stems from the lattice mirror symmetry that precludes Landau-level mixing. Introducing a non-zero displacement field \(D\) disrupts this mirror symmetry, facilitating the hybridization between the monolayer-like and bilayer-like Landau levels.  This inter-band coupling enhances the Landau level mixing factor $\eta$ and activates three-body interactions -- both of which explicitly break the PHS of FQHs. As a result, conventional FQHs are completely destabilized, offering a route to engineer symmetry breaking of FQHs in a controlled way. We establish that the PHS breaking in TLG is of extrinsic origin and is fundamentally distinct from the intrinsic, interaction-driven symmetry breaking observed in the lowest Landau levels of single-layer and bilayer graphene.
\end{abstract}

\maketitle


{\it Introduction} -- The Fractional Quantum Hall (FQH) effect, observed at ultra-low temperatures in minimally-disordered two-dimensional semiconductors subjected to a perpendicular magnetic field $B$, is characterized by simultaneous vanishingly small longitudinal conductance $G_{xx}$ and quantized transverse conductance $G_{xy} = \nu e^2/h$~\cite{PhysRevLett.48.1559, RevModPhys.71.S298}.  The parameter $\nu =n/n_\phi$, a rational fraction,  is the topological invariant of the FQH phase~\cite{PhysRevLett.71.3697}. Here, $n$ is the areal charge carrier density, and $n_\phi$ is the areal density of flux quanta threading the material. Several characteristics of this correlated incompressible FQH fluid phase can be understood through a framework that transforms the highly interacting electrons (by associating with them an even number of vortices) into weakly-interacting excitations, known as composite fermions (CF)~\cite{PhysRevB.47.7312, PhysRevLett.63.199, PhysRevX.5.031027}. The CF experience an `effective' magnetic field $B_{eff} = B/(2p\pm1)$, with $p = \nu/(1-2\nu) = \pm1,\pm2,\pm3...$ .
Away from half-filling of the Landau levels, this process effectively maps the $\nu^{\mathrm{th}}$ FQH state of electrons into the $p^{\mathrm{th}}$ integer quantum Hall (IQH) state of the CF~\cite{PhysRevB.47.7312, PhysRevLett.71.3846, PhysRevLett.71.3850,2023arXiv231206194K}.

A fundamental property of FQH phases is particle-hole (PH) symmetry~\cite{PhysRevB.54.R17323, PhysRevB.92.165125, PhysRevB.55.15552, PhysRevB.96.245142}. Without Landau level (LL) mixing, the Hamiltonian of the two-dimensional electron gas can be projected into a single LL; the other LLs can be neglected as being completely full or empty. The extent of LL mixing is parameterized by  $\eta = E_c/E_{cyc}$. $E_c=e^2/4\pi\epsilon\epsilon_0 l_B$ is the Coulomb interaction energy scale, $l_B = \sqrt{\hbar/(eB)}$ is the magnetic length, $\epsilon$ is the dielectric constant of the medium and $E_{cyc}$ is the cyclotron energy gap~\cite{PhysRevB.80.121302, PhysRevB.87.245129, PhysRevB.87.245425, PhysRevB.47.7312, PhysRevX.5.031027}. $\eta$ is a measure of the probability that electrons in a partially occupied LL can make a virtual hop to adjacent LLs. Single-LL projection is quantitatively accurate if $\eta <2$. In such a system, the Hamiltonian for FQH states $\nu = N + p/(2p+1)$ is related to those for $\nu = N+1-p/(2p+1)$ by an anti-unitary transformation; the $\nu = N + p/q$ state and its hole-conjugate state $\nu = N+1-p/q$ have identical energy scales~\cite{PhysRevB.29.6012, PhysRevLett.118.206602,PhysRevB.96.245142}.  For materials characterized by a parabolic energy dispersion (e.g., semiconductor quantum wells, bilayer graphene), $\eta \propto B^{-1/2}$~\cite{PhysRevB.80.121302, PhysRevB.87.245129, PhysRevB.87.245425, PhysRevB.47.7312, PhysRevX.5.031027}. Conversely, for systems with a linear dispersion (like single-layer graphene), $\eta = e^2/(\epsilon v_f \hbar)$ is independent of $B$ ($v_F$ is the Fermi velocity); one can tune $\eta$ instead by varying $\epsilon$~\cite{PhysRevLett.113.086401}. Despite its obvious significance to a complete understanding of FQH in graphene, experimental studies of the effect of a \textit{controlled} enhancement of LL mixing on PH symmetry in the FQH regime are lacking.

In this article, we establish the PH symmetry of FQH states around half-filling in a system characterized by linear energy dispersion and then demonstrate its controlled violation through an \textit{external} parameter.
Since electron-electron interactions are PH symmetric, the tuning parameter must be band structure or chemical potential~\cite{PhysRevB.96.075148}. We choose ABA-stacked trilayer graphene (TLG) as the material platform as its inter-LL spacing and, consequently, $\eta$ can be precisely controlled by an external displacement field $D$ applied perpendicular to its plane~\cite{PhysRevB.87.115422, PhysRevB.101.245411, PhysRevLett.121.167601,doi:10.1021/acs.nanolett.2c00435, PhysRevLett.117.066601} (Supplementary figure 4). Our experiments reveal PH symmetry of FQH states around half-filling when inter-LL spacing is large ($\eta$ is small), and its controlled violation as $D$ reduces the inter-LL spacing (leading to enhanced $\eta$). Calculations of the LL spectrum versus interlayer potential $\Delta_1$, corroborated by direct measurements of $\eta$,  confirm that PH asymmetry in TLG over a specific range of $D$ stems from inter-band Landau level mixing, which amplifies the role of three-body interactions -- a factor known to explicitly break PH symmetry \cite{PhysRevB.96.245142,PhysRevLett.113.086401,PhysRevB.80.121302,PhysRevB.104.L081407}.

\begin{figure}[t]
	\includegraphics[width=\columnwidth]{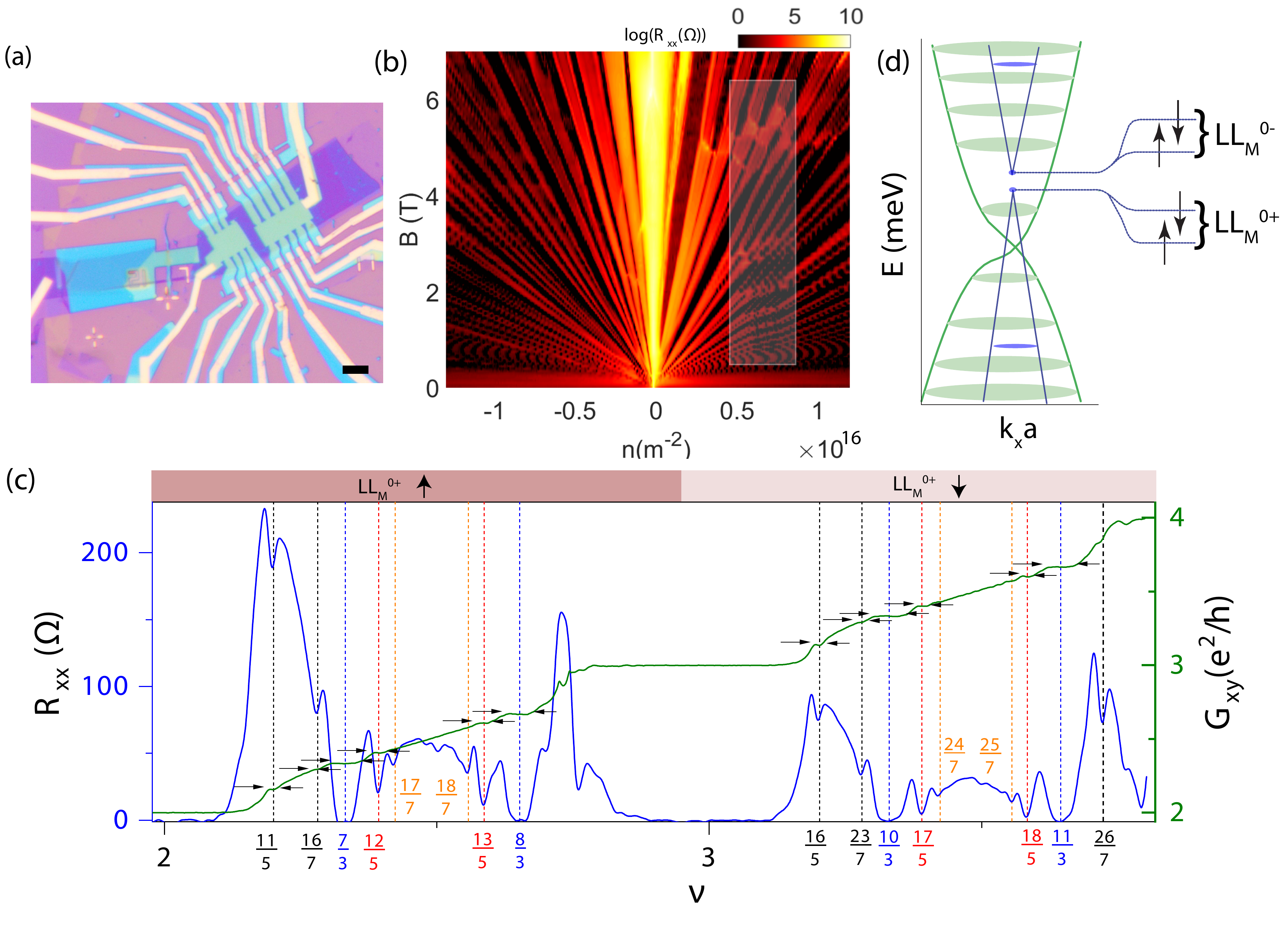}
	\small{\caption{ \textbf{Device characterization and Fractional Quantum Hall states in ABA TLG}. (a) Optical microscope image of the device1. The scale bar is  $5~\mathrm{\mu m}$. (b) 2D map of longitudinal conductance $R_{xx}$ in the $B$ and number density $n$ plane. The rectangular box marks the crossings between the four $LL_M^{0}$ LLs with the BLL LLs. (c) Plots of $R_{xx}$ (left y-axis) and $G_{xy}$ (right y-axis) as a function of filling factor $\nu$ at $B =12$ T and $T = 0.02$ K. (d) A schematic band structure of ABA trilayer graphene in the $E-k$ space at $D=0$~V/nm. The green ellipses represent bilayer-like LLs. The purple discs represent the monolayer-like LLs. The notation $LL_M^{\alpha\pm}$ denotes that the LL is an eigenstate of the monolayer part of the Hamiltonian, $\alpha$ is the LL index in the MLL band, `$+$' and `$-$' represent the $K$ and $K'$ valleys respectively. The spin degenerate $LL_M^{0+}$ LL resides at the maxima of the MLL valence band while the $LL_M^{0-}$ is at the minima of the MLL conduction band. The data are from device~1.}
		\label{fig:fig1}}
\end{figure}

ABA trilayer graphene flakes (hereafter referred to as TLG), mechanically exfoliated on \ch{SiO2}, were identified through optical contrast and Raman spectroscopy~\cite{cong2011raman}.   TLG devices, encapsulated between single-crystalline hexagonal boron nitride (hBN)  (Fig.~\ref{fig:fig1}(a)), were prepared using the dry-transfer technique~\cite{doi:10.1126/science.1244358, doi:10.1021/acs.nanolett.3c00045, PhysRevLett.129.186802, doi:10.1021/acs.nanolett.3c04223, Pizzocchero2016}. Dual graphite electrostatic gates were used to simultaneously tune the number density $n=[(C_{tg}V_{tg}+C_{bg}V_{bg})/e]$ and the displacement field $D=(C_{bg}V_{bg}-C_{tg}V_{tg})/{2\epsilon_0}$. Here $C_{bg}$ ($C_{tg}$) is the back-gate (top-gate) capacitance, $V_{bg}$ ($V_{tg}$) is the back-gate (top-gate) voltage and $\epsilon_0$ is the vacuum permittivity.  The distinct BLL and MLL bands are apparent from the Landau fan diagram in Fig.~\ref{fig:fig1}(b); the data were measured at a temperature $T =0.02$~K and $D =0$~V/nm. Multiple LL crossings (marked by the shaded region) originate due to zeroth-LL $LL_{M}^{0}$ of the monolayer-like band crossing several bilayer-like LLs \cite{shimazaki2016landau}. We present the data from two different devices having identical architecture in this Letter; \textbf{Device~1} had mobility $\sim 620,000$~$\mathrm{cm^2V^{-1}s^{-1}}$ while \textbf{Device~2} had mobility  $\sim 1,100,000$~$\mathrm{cm^2V^{-1}s^{-1}}$. Data from another device is presented in Supplemental Materials.

Fig \ref{fig:fig1}(c) is the plot of longitudinal conductance $R_{xx}$ and transverse conductance $G_{xy}$ versus the filling factor measured at a fixed perpendicular magnetic field, $B =12$~T, $T =0.02$~K and $D =0.1$~V/nm. The dips in $R_{xx}$ and plateaus in $G_{xy}$ identify the FQH states at the corresponding filling factors marked along the x-axis. These FQHs originate from the $LL_{M}^{0+}$ LL (Fig \ref{fig:fig1}(d)). We also find indications of the 4-flux quanta CF states ($\nu = 16/5, 23/7$ and $26/7$), observed previously in monolayer graphene through quantum capacitance~\cite{PhysRevLett.122.137701} and compressibility measurements~\cite{PhysRevLett.111.076802}.

\begin{figure}[t]
	\includegraphics[width=\columnwidth]{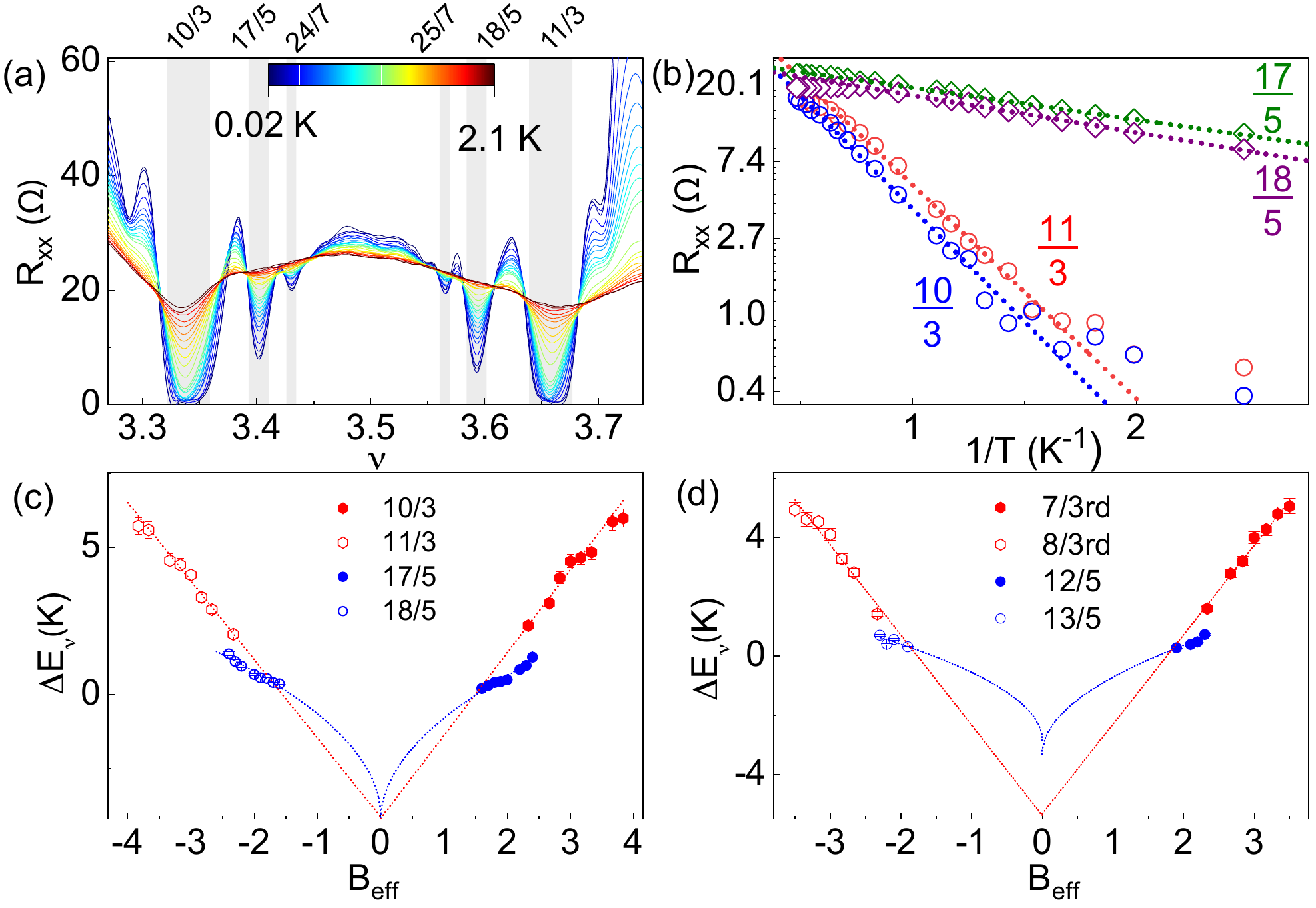}
	\small{\caption{ \textbf{Particle hole symmetry and activation gaps of FQHs}. (a) Plots of $R_{xx}$ versus filling factor $\nu$ at a few representative $T$ for $D = 0.1$~V/nm between $\nu =3$ and $\nu = 4$. (b) Symbols are the Arrhenius plots for various FQH states between $\nu =3$ and $4$. The dashed lines are the fit to data points using Arrhenius fit. Plot of activation gaps as a function of  $B_{eff}$ for FQHs between (c) $\nu =3$ and $4$, and (d) $\nu =2$ and $3$. The red dashed lines are linear fits to the data points. The dashed blue lines are $\sqrt{B}$ fits to the data points. The data are from device~1.}
		\label{fig:fig2}}
\end{figure}

\textit{PHS at small values of \texorpdfstring{$D$}{D}} -- Fig.~\ref{fig:fig2}(a) shows plots of the longitudinal resistance $R_{xx}$ as a function of the filling factor $\nu$ at $B =11$~T for several representative temperatures. The regions of vanishing $R_{xx}$ arise from the formation of the FQH incompressible state $\nu = 10/3, 17/5$, and $24/7$  (equivalently $p = 1,2,3$), and their hole-conjugate states  $11/3, 18/5$ , and $25/7$. With increasing $p$, the temperature range over which the FQH states are discernible shrinks, consistent with the predicted dependence of the FQH gap on $p$. The disappearance of the particle and hole states at very similar temperatures suggests the presence of PH symmetry around $\nu = 3+1/2$.

We quantify the PH symmetry through the activation energy gaps extracted using the relation  $R_{xx} \propto \mathrm{exp}(-\Delta E_\nu/2k_BT)$. Several of these FQH states persist to relatively low magnetic fields ($\sim 6$~T), providing a sufficient range to probe the dependence of activation gaps on the magnetic field. Representative plots of the Arrhenius fits are shown in Fig \ref{fig:fig2}(b) (in Supplementary Note 2 for other $B$ values). Plots of $\Delta E_\nu$ versus $B_{eff}$ for FQH transitions between $\nu =3$ and $4$ and between $\nu =2$ and $3$ are shown in Fig \ref{fig:fig2}(c) and Fig \ref{fig:fig2}(d), respectively. The measured gaps exhibit a reflection symmetry around $B_{eff} = 0$, confirming PHS about half-filling. The gaps for $\nu = 12/5$, $13/5$, $17/5$ and $18/5$ have a $\sqrt{B}$-dependence (dashed blue lines in Fig \ref{fig:fig2}(c-d)). For spinless CF, the energy gaps separating CF LLs originate from Coulomb interactions~\cite{PhysRevLett.92.156401, PhysRevLett.122.137701} and  are expected to have a $\sqrt{B}$-dependence; $\Delta E_\nu = \hbar eB_{eff}/m_{eff} - \Gamma$. Here, $\Gamma$ is the disorder-induced Landau level broadening and $m_{eff}=\alpha m_e\sqrt{(2p+1)B_{eff}}$ is the effective CF mass  ($\alpha$ is the effective mass parameter and $m_e$ the mass of free electrons).

In contrast, the $B$-dependence of the gaps for $\nu = 7/3$,  $8/3$, $\nu = 10/3$ and  $11/3$ can be fitted equally well with either a linear ${B}$-dependence (dotted red lines in Fig \ref{fig:fig2}(c-d)) or a $\sqrt{B}$-dependence (Supplementary Note 11). A linear-$B$ dependence of the energy gap is observed if it is dominated by the  Zeeman energy $(\propto B)$ rather than the cyclotron energy $(\propto \sqrt{B})$~\cite{https://doi.org/10.1002/andp.20025141205, PhysRevB.74.165325}. In this case, one quantifies these gaps using  $\Delta E_\nu=\frac{1}{2}\mu_B\mathbf{g}(2p+1)B_{eff}-\Gamma$ with $\mathbf{g}$ the effective Land\'e g-factor. Given this ambiguity of the fits, the origin of the gap in the third and two-thirds states is uncertain. The CF effective mass parameter, Land\'e g-factor, and the disorder broadening extracted from the fits show a remarkable PH-symmetry (Supplementary Note~13).

\begin{figure}[t]
	\includegraphics[width=\columnwidth]{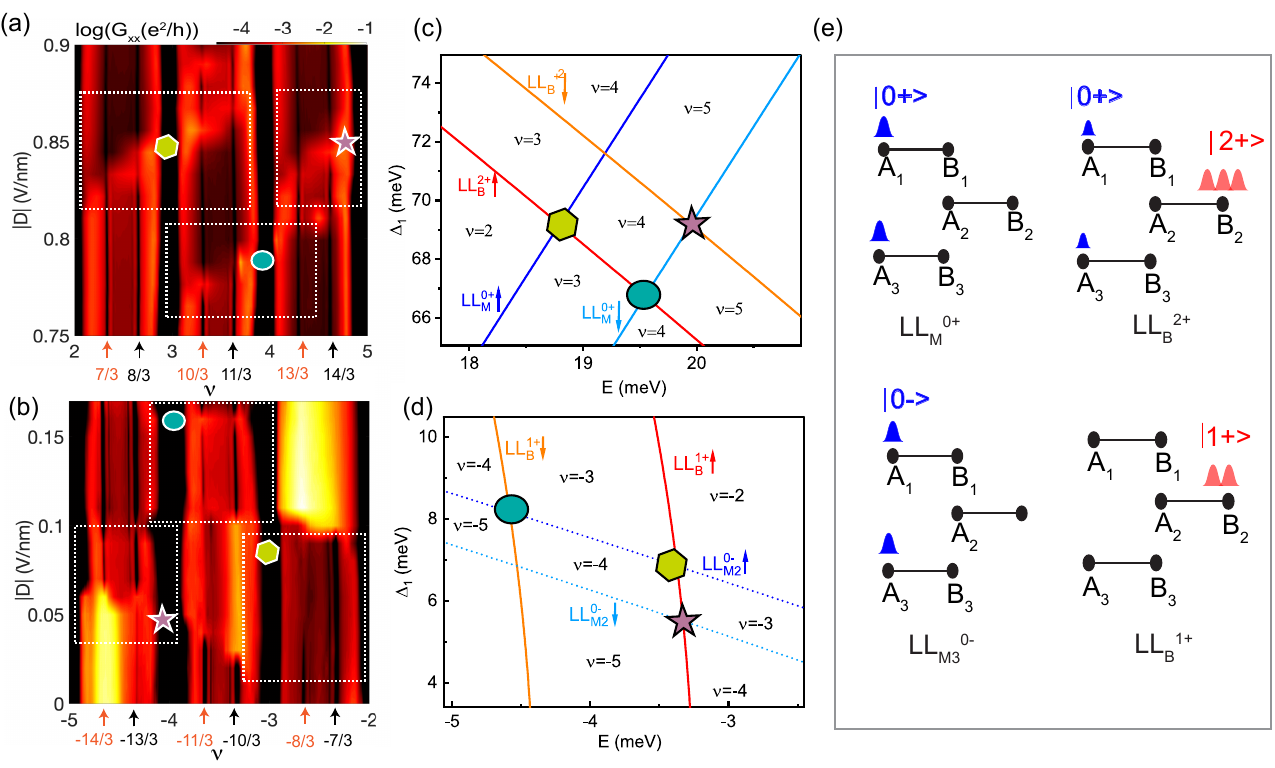}
	\small{\caption{ \textbf{Effect of Landau level mixing on FQHs in ABA trilayer graphene} Contour plots of $G_{xx}$ as a function of $D$ and  $\nu$ for (a) electron doping, and (b) hole doping. The data are taken at $B= 10$~T. The white dotted rectangles mark the regions of PHS violation. Simulated Landau level spectrum as a function of interlayer potential $\Delta_1$ and energy $E$ for (b) electron doping and (e) hole doping. The data are plotted over the same range $\Delta_1$ ($D$) as in (a) and (b), respectively. The filled symbols indicate LL crossings, which are also marked in the experimental data. The data are from device~2. (e-f) Schematics of the wavefunctions localized on different atomic sites in TLG for the different LLs. In the notation $\ket{N,\xi}$, $N$ denotes the orbital index, and $\xi$ is the valley index. }
		\label{fig:Fig3}}
\end{figure}

\textit{Violation of PHS in the vicinity of Landau level mixing} -- Next, we discuss the effect of a finite $D$ field on the FQHs. We plot the contour plot of $G_{xx}$ as a function of $D$ and $\nu$ in the range $\nu=2$ to $5$  in Fig.~\ref{fig:Fig3}(a) and for $\nu=-2$ to $-5$  in Fig.~\ref{fig:Fig3}(b). The dashed rectangles highlight the regions where either the FQHS or its hole-conjugate state vanishes, indicating a strong violation of PHS over values of $D$ that are distinct for each LL. Supplementary Table~1 provides a comprehensive summary of PH symmetry violation and lists the FQHS most affected by it in various filling factors $\nu$.

To understand the origin of this PH asymmetry over specific $D$ ranges, we simulated the LL spectrum using single particle tight binding calculations based on the Slonczewski-Weiss-McClure model~\cite{PhysRevB.87.115422, PhysRevLett.121.167601} with varying potential difference $\Delta_1$ between the two outer layers. For TLG
$\Delta_1 = [(d_{\perp}/2\epsilon_{TLG})\times~D]e$, leading to $\Delta_1 (\mathrm{meV}) = 82~D~(\mathrm{V/nm})$~\cite{PhysRevLett.132.096301,PhysRevLett.121.167601}. Here, $d_{\perp}=0.67$~nm is the separation between the top and bottom layers of TLG, $\epsilon_{TLG}$ is the dielectric constant of the TLG, and $e$ is the electronic charge. We use the following tight-binding parameters: $\gamma_0=3.1~\mathrm{eV},\,\gamma_1=0.39~\mathrm{eV},\,\gamma_2=-0.005 ~\mathrm{eV},\,\gamma_3=0.275~\mathrm{eV},\,\gamma_4=0.041 ~\mathrm{eV},\,\gamma_5=0.005 ~\mathrm{eV},\,\delta=0.0108~\mathrm{eV},\,\textrm{and}\,\Delta_2=0.003 ~\mathrm{eV}$~\cite{PhysRevLett.121.167601}.
The resultant Landau spectrum is shown in Fig.~\ref{fig:Fig3}(c) for $\nu=2$ to $5$ and in  Fig.~\ref{fig:Fig3}(d) for $\nu=-2$ to $-5$. Applying $D$ negates the mirror symmetry in pristine TLG and induces mixing between LLs from monolayer-like and bilayer-like bands. We identify multiple crossings between spin-split LLs of $LL_M^{0+}$ and $LL_B^{2+}$ (Fig \ref{fig:Fig3}(c)) and between spin-split LLs of $LL_{M2}^{0-}$ and $LL_B^{1+}$ (Fig \ref{fig:Fig3}(d)). Each filled symbol in Fig.~\ref{fig:Fig3}(c-d) denotes a specific LL crossing, with its counterpart marked in the experimental data in Fig.~\ref{fig:Fig3}(a-b). We find a one-on-one correspondence between the regions of PH asymmetry in Fig.~\ref{fig:Fig3}(a-b) with the regions of LL crossing in Fig.~\ref{fig:Fig3}(c-d).

\begin{figure}[h]
	\includegraphics[width=\columnwidth]{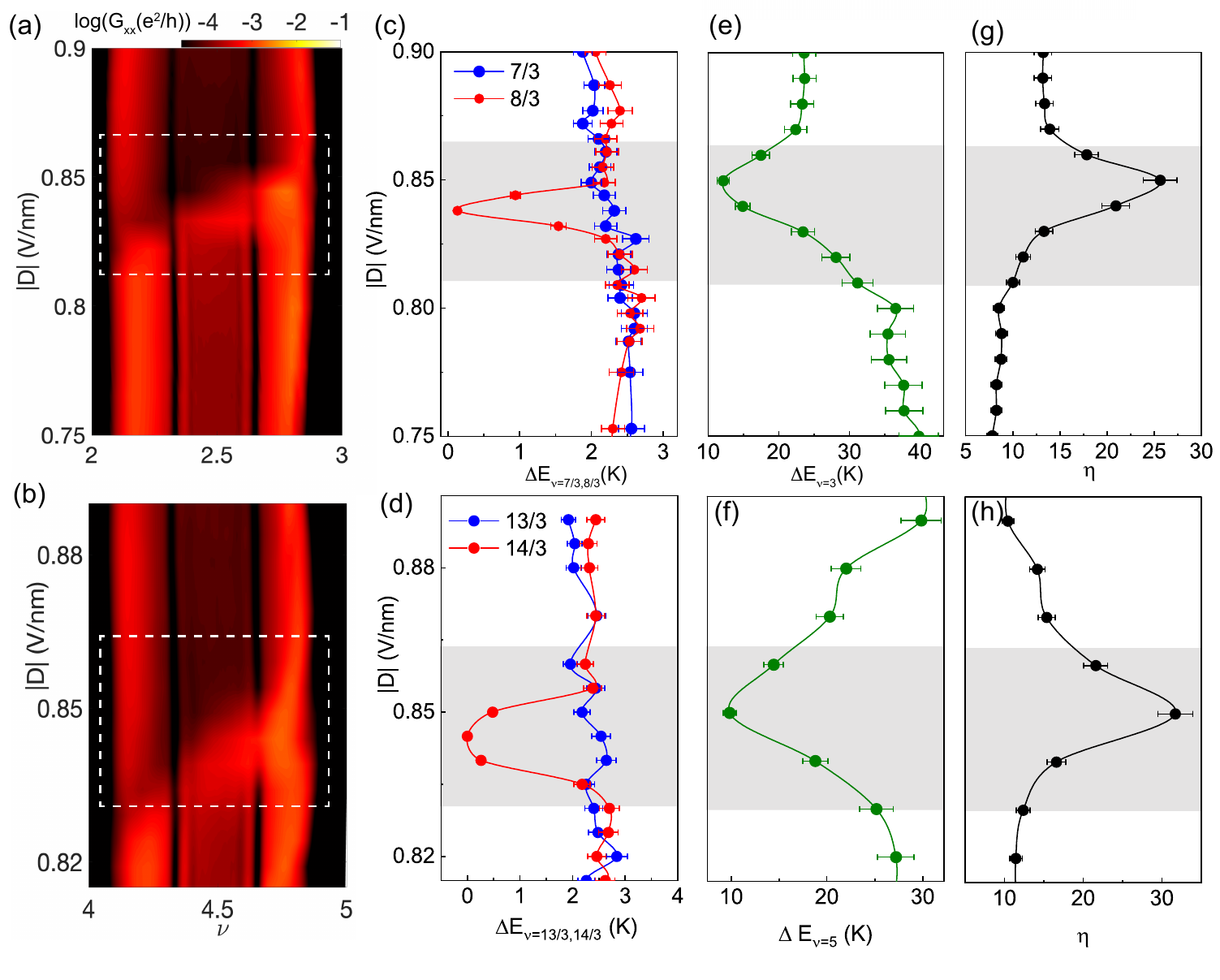}\caption{\textbf{Controlled violation of PHS.} Contour plot of $G_{xx}$ as a function of $D$ and $\nu$ around the LL crossing between filling factor (a) $\nu=2$ and $3$, (b) $4$ and $5$. The white dotted rectangle marks the region of observed PHS violation. Plot of $D$-dependence of activation gap $\Delta E$ for (c) $\nu=7/3$ (blue filled circles) and $8/3$ (red filled circles), (d) $\nu=13/3$ (blue filled circles) and $14/3$ (red filled circles). The gray-shaded region demarcates the region of PHS violation. Plot of $D$-dependence of activation gap for (e) $\Delta E_{\nu=3}$, (f) $\Delta E_{\nu=5}$. $\Delta E$ has a sharp minimum in the gray-shaded region. (g) ,(h) Plot of $D$-dependence of Landau level mixing parameter $\eta$ calculated from the data in (e) and (f). The gray-shaded region marks the region of enhanced $\eta$ due to the coming together of $LL_M^{0+}\uparrow$ and the $LL_B^{2+}\uparrow$ Landau levels for (a) and $LL_M^{0+}\downarrow$ and the $LL_B^{2+}\downarrow$ for (b). The data are from device~2.}
	\label{fig:fig4}
\end{figure}

Having established a definite relation between PHS violation and LL crossing, we focus on the $\nu=3$,$5$ and quantify the extent of PH asymmetry due to the crossing of $LL_M^{0+}$ and $LL_B^{2+}$ levels (marked by the filled pentagon and star in Fig.~\ref{fig:Fig3}(c)). Fig. \ref{fig:fig4}(a) and (b) shows a high-resolution map of measured $G_{xx}$ around the LL crossing point. The dashed rectangle marks the region where hole-conjugate FQHs states ($\nu=8/3$ and $13/5$ in Fig \ref{fig:fig4}(a) and $\nu=14/3,23/5$ in Fig \ref{fig:fig4}(b)) vanish.  The activation gaps of $\nu=7/3$ and $8/3$ ($\nu=13/3$ and $14/3$) as a function of $D$ are shown in Fig.~\ref{fig:fig4}(c) (Fig \ref{fig:fig4}(d)). For $|D|<0.82$~V/nm, $\Delta E_{7/3} \simeq \Delta E_{8/3}$, indicating the   PH symmetry of FQHs. This is consistent with the calculated LL spectrum -- the large inter-LL spacing between $LL_M^{0+}\uparrow$ and $LL_B^{2+}\uparrow$ in this range of $D$ values prevents perturbation to the FQHs residing in $LL_M^{0+}\uparrow$. However, for $0.82<|D|<0.86$~V/nm (gray-shaded region in Fig.~\ref{fig:fig4}(c)) the distinction between the $D$-response of the $\nu=7/3$ and $8/3$ is remarkable. $\Delta E_{7/3}$ (blue closed symbols) remains almost unchanged at all $D$. In contrast, $\Delta E_{8/3}$ (marked by blue closed circles) undergoes a sharp minimum. A comparison with the simulated LL spectrum (Fig.~\ref{fig:Fig3}(c)) reveals that this is the regime of LL crossing. With increasing $D$, the inter-LL spacing again increases, restoring PH symmetry for $|D|>0.86$~V/nm. Similar trend is seen for $\Delta E_{13/3,14/3}$ (Fig. \ref{fig:fig4}(d)).

Furthermore, we measured the excitation gap of the $\nu=3$ and $5$ integer quantum Hall phase as a function of $D$ where $\Delta E_{3} = E(LL_{B}^{2+}\uparrow)-E(LL_{M}^{0+}\uparrow)$ and $\Delta E_{5} = E(LL_{B}^{2+}\downarrow)-E(LL_{M}^{0+}\downarrow)$  -- the data are plotted in Fig.~\ref{fig:fig4}(e) and \ref{fig:fig4}(f). We observe that within the range of $D$ where PH symmetry is violated (gray-shaded region), the value $\Delta E_{3,5}$ falls by a factor of $2.5$, indicating a sharp reduction in inter-LL spacing. Note that our single-particle picture predicts a crossing between these LLs; in reality, any interaction will lift this accidental degeneracy, leading to a reduced but finite activation gap at this point.
From the measured $\Delta E_{3,5}$, we estimate $\eta=E_c/\Delta E_{3,5}$. The results plotted in Fig.~\ref{fig:fig4}(g) and Fig.~\ref{fig:fig4}(h) show that $\eta(D)$ has a maximum in the vicinity of the PHS violation (gray-shaded region). Similar quantitative analyses for other FQH states are presented in Supplementary Section 8.

The simultaneous occurrence of pronounced PHS violation and a sharp dip in the activation gap of the underlying IQH state (equivalently, a marked enhancement of the Landau level mixing parameter $\eta$)  within the same narrow window of external displacement field is striking. The measurements were repeated for several different values of $B$; at each magnetic field, the range of $D$ over which PH symmetry is violated matches the $\Delta_1$ values where LL overlap is theoretically expected (Supplementary section 5). This concurrence provides compelling evidence of a strong and direct causal link between $D$-induced modifications to the Landau level spectrum~\cite{PhysRevLett.117.066601, PhysRevB.87.115422, PhysRevLett.121.056801} and the observed breakdown of PHS in the system~\cite{PhysRevLett.113.086401}.

\textit{Discussion}-- The zeroth LL of monolayer graphene is PH symmetric. Any observed asymmetry typically stems from valley isospin transitions between spin split LLs of $LL_{M}^{0+}$ and $LL_{M}^{0-}$, which stabilize even-denominator FQHs at $\nu=\pm 1/2$ and lead to slight differences in the excitation gaps of particle and hole-conjugate states \cite{PhysRevLett.121.226801,Zibrov2018}. In bilayer graphene, the particle-hole symmetry breaking is attributed to the interaction-induced breaking of intrinsic symmetries of the system, giving rise to elegant cascaded of iso-spin transitions within the multi-component zeroth LL \cite{Zibrov2017,PhysRevX.12.031019,doi:10.1126/science.aao2521,kumar2024quarterhalffilledquantumhall}. This manifests in the formation of Pfaffian phase daughter state at $\nu= 1+7/13$ and $\nu=1+3/5$ FQHs near the isospin phase transition \cite{Zibrov2017}.

However, the situation in TLG is significantly different: the wavefunction of the carriers in the $LL_{M}^{0+}$ is single particle:$\ket{\psi_M0^+} = \ket{0}\circledast (\ket{A1}-\ket{A3})/\sqrt{2}\circledast\ket{K}$~\cite{shimazaki2016landau} (Fig \ref{fig:Fig3}(e)) with minimal contribution from valley isospin transitions even at $D=0$ (see Supplementary figure 4(c)). Increasing $D$ causes the $LL_{M}^{0+}$ and $LL_{M}^{0-}$ bands to diverge from each other, further ruling out the possibility of $D$ induced valley isospin transitions as an agent for PHS breaking. The preceding discussion establishes that understanding the violation of PHS at $D\neq0$ in TLG does not necessitate interactions or related symmetry breaking: instead it is a case of explicitly broken PHS due to virtual scattering between $LL_{M}^{0+}$ and $LL_{B}^{2+}$ LL. Such inter-orbital mixing enhances the role of three body interactions which are predicted to completely destabilizing the conventional FQHs. Supplementary Table 1 shows the complete summary of PH symmetry violation and the states most effect in different filling factors $\nu$ due to mixing between BLL and MLL LLs. While interaction-driven symmetry breaking may enhance particle-hole asymmetry in TLG at a finite $D$, our study establishes that it is not the primary factor driving this effect. This point distinguishes the PHS violation seen in Bernal stacked trilayer graphene from previous studies in other forms of graphene.

Note that unlike in trilayer graphene, LL spacing between ZLL and higher LLs in SLG and BLG is largely fixed by the bandstructure and
is only weakly tunable with $D$, making it difficult to probe such Landau level mixing effects \cite{Zibrov2017,PhysRevB.91.115405}.

\textit{Conclusion} -- To summarize, we present evidence of PHS around half-filling in a multiband system, viz. ABA trilayer graphene. For low $|D|$, the MLL and BLL LL are mutually non-interacting due to the lattice mirror symmetry. This precludes LL mixing, and the FQH states are PH symmetric. With increasing $|D|$, two things happen simultaneously: (1) the mirror symmetry breaks, and (2) the $LL_M^{0+}$  and the $LL_B^{2+}$ LLs levels approach each other. At sufficiently large values of \( |D| \) (near LL crossing between MLL and BLL LLs), the requirement for particle-hole symmetry -- specifically, that the cyclotron energy is significantly greater than the interaction strength -- ceases to hold. This condition is characterized by enhancing the Landau level mixing parameter \( \eta \). The coming together of the LLs enhances the rate of virtual excitations between them, eventually destroying the PH symmetry around a critical displacement field,  $|D_C| = 0.82$~V/nm. Increasing $D$ further moves the LLs apart, reducing $\eta$ and restoring the PH symmetry.

\textit{Acknowledgments} -- The authors acknowledge fruitful discussions with Yuval Gefen. A.B. acknowledges funding from the U.S. Army DEVCOM Indo-Pacific (Project number: FA5209   22P0166). K.W. and T.T. acknowledge support from the JSPS KAKENHI (Grant Numbers 21H05233 and 23H02052) and World Premier International Research Center Initiative (WPI), MEXT, Japan.

\clearpage

\clearpage

\section*{ Supplementary Information }
\addto\captionsenglish{\renewcommand{\tablename}{}}
\renewcommand{\theequation}{Supplementary Equation~\arabic{equation}}
\renewcommand{\thesection}{S~\arabic{section}}
\renewcommand{\thefigure}{{Fig.~\arabic{table}}}
\renewcommand{\thetable}{{Supplementary Table~\arabic{table}}}

\setcounter{section}{0}
\setcounter{table}{0}

\section{Device fabrication, schematics and characterization}
Bernal-stacked trilayer graphene (TLG), hexagonal boron nitride (hBN), and graphite flakes were obtained by mechanical exfoliation of single crystals. ABA TLG flakes are identified from optical contrast and Raman spectroscopy \cite{cong2011raman}. A dry pickup and transfer technique \cite{pizzocchero2016hot} was used to fabricate the graphite-encapsulated hBN/ABA-TLG/hBN heterostructures.  The sequence of layers is shown schematically in Fig~\ref{fig:figS1}(a). 1-D metallic contacts were defined by e-beam lithography, followed by reactive ion etching using $\mathrm{CHF_3/O_2}$ gas and deposition of Cr/Pd/Au (3~nm/12~nm/55~nm) metallic contacts \cite{doi:10.1126/science.1244358}. The device was finally etched into a Hall bar shape. We use graphite as the bottom and top gate electrodes with hBN as the gate dielectric. A Si/SiO$_2$ back gate is used to dope the graphene contacts to avoid the formation of p-n junctions. Dual gate configuration is used to simultaneously tune the vertical displacement field $D=[(C_{bg}V_{bg}-C_{tg}V_{tg})/2\epsilon_0]$ and areal carrier density  $n=[(C_{bg}V_{bg}+C_{tg}V_{tg})/e]$ across the sample independently. Here $C_{bg}$ ($C_{tg}$) is the back-gate (top-gate) capacitance, and $V_{bg}$ ($V_{tg}$) is the back-gate (top-gate) voltage. Fig.~\ref{fig:figS1}(b) shows a schematic of the device with the different gates.

A standard low-frequency lock-in detection technique is used to perform electrical transport measurements with the current $i=10$~nA and at frequency $13$~Hz. Fig.~\ref{fig:figS1}(c) shows the longitudinal resistance $R_{xx}$ versus $V_{bg}$ of the device1 measured at $B=0$ and $T=20$~mK. The measured $R_{xx}$ is fitted with the equation~\cite{WOS:000292115900140}:
\begin{equation}
	R_{xx} = \frac{L}{We\mu \sqrt{n_0^2+(\frac{C_{bg}(V_{bg}-V_{dp})}{e})^2}}.
	\label{Eqn:rxx}
\end{equation}
where $L$ is the distance between two voltage probes, $W$ is width of the sample, and $\mu$ is the mobility of the sample. $V_{dp}$ is the voltage value at maximum resistance. Fits of the data using~\ref{Eqn:rxx} gives  $\mu= 620,000$~$\mathrm{cm^2V^{-1}s^{-1}}$ and impurity carrier density $n_0 = 7.81\times10^{9}$~$\mathrm{cm^{-2}}$ demonstrating the high quality of the sample. The corresponding numbers for the device2: $\mu= 11,00,000$~$\mathrm{cm^2V^{-1}s^{-1}}$ and $n_0 = 4\times10^{9}$~$\mathrm{cm^{-2}}$ and for the device3 are $\mu= 320,000$~$\mathrm{cm^2V^{-1}s^{-1}}$ and $n_0 = 1.42\times10^{9}$~$\mathrm{cm^{-2}}$.
\begin{suppfigure}[b]
	\includegraphics[width=0.9\columnwidth]{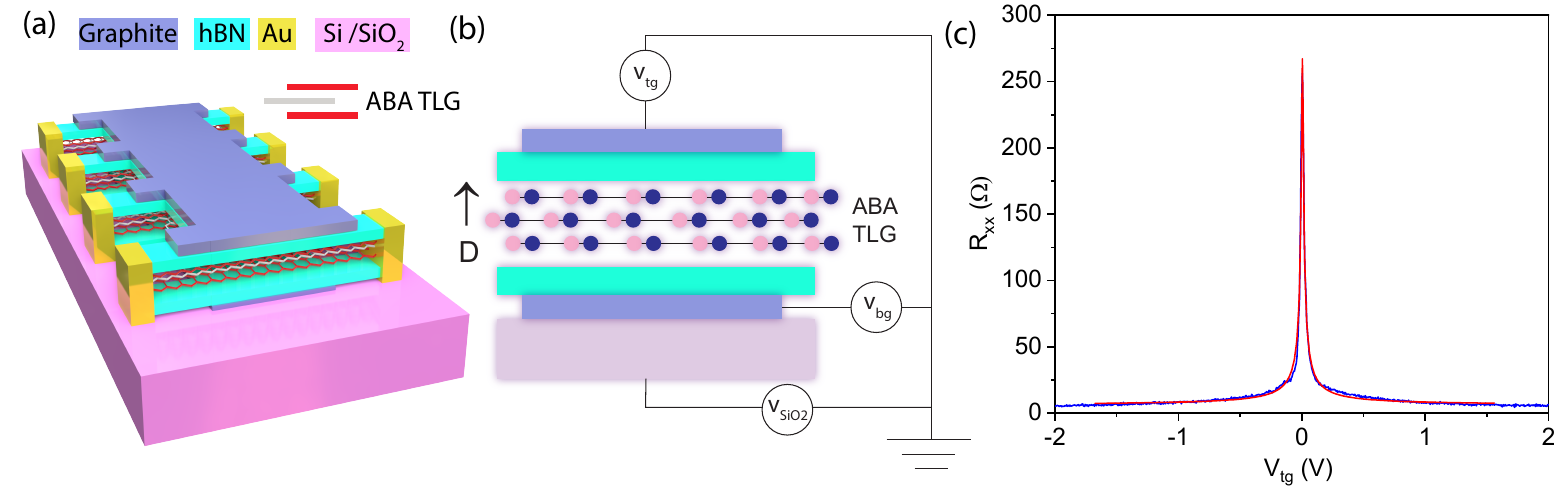}
	\caption{\textbf{Device schematic and characteristics:} (a) Device schematic of the ABA TLG encapsulated between two hBN flakes. (b)  Schematic of the device showing the different gate configurations used in the measurements. The top and bottom graphite gates were used to tune the vertical displacement field $D$ and number density $n$ across the sample. A Si/SiO$_2$ back gate was used to dope the contacts.    (c) The solid line plots longitudinal resistance as a function of top-gate voltage measured at $B= 0$~T and $T = 0.02$~K for device1. The solid red line is the fit of the data to ~\ref{Eqn:rxx}.}
	\label{fig:figS1}
\end{suppfigure}

\section{Activated behavior of \texorpdfstring{$R_{xx}$}{Rxx} in FQH regime.}

The activation gaps at the $R_{xx}$ minima are extracted from the relation \cite{PhysRevB.106.L041301,PhysRevLett.55.1606,PhysRevLett.70.2944,PhysRevLett.124.156801}:
\begin{equation}
	R_{xx} = R_0 \mathrm{exp}(-\Delta E_\nu/2k_BT).
	\label{Eqn:arrehniuseq}
\end{equation}
Fig.~\ref{fig:figS2}(a-c) show the plots of longitudinal resistance as a function of filling factor $\nu$ at $B=10$ T, $8.5$ T, and $7$~T, respectively, for the FQH states around $\nu = 7/2$; the data are for the device~1. Fig.~\ref{fig:figS2}(d), (e), and (f) show the corresponding activated fits to the data points extracted from the minima in $R_{xx}$. From the slopes of these plots, we extract the values of $\Delta E_\nu$ using Eqn.~\ref{Eqn:arrehniuseq}.

\begin{suppfigure}
	\includegraphics[width=0.75\columnwidth]{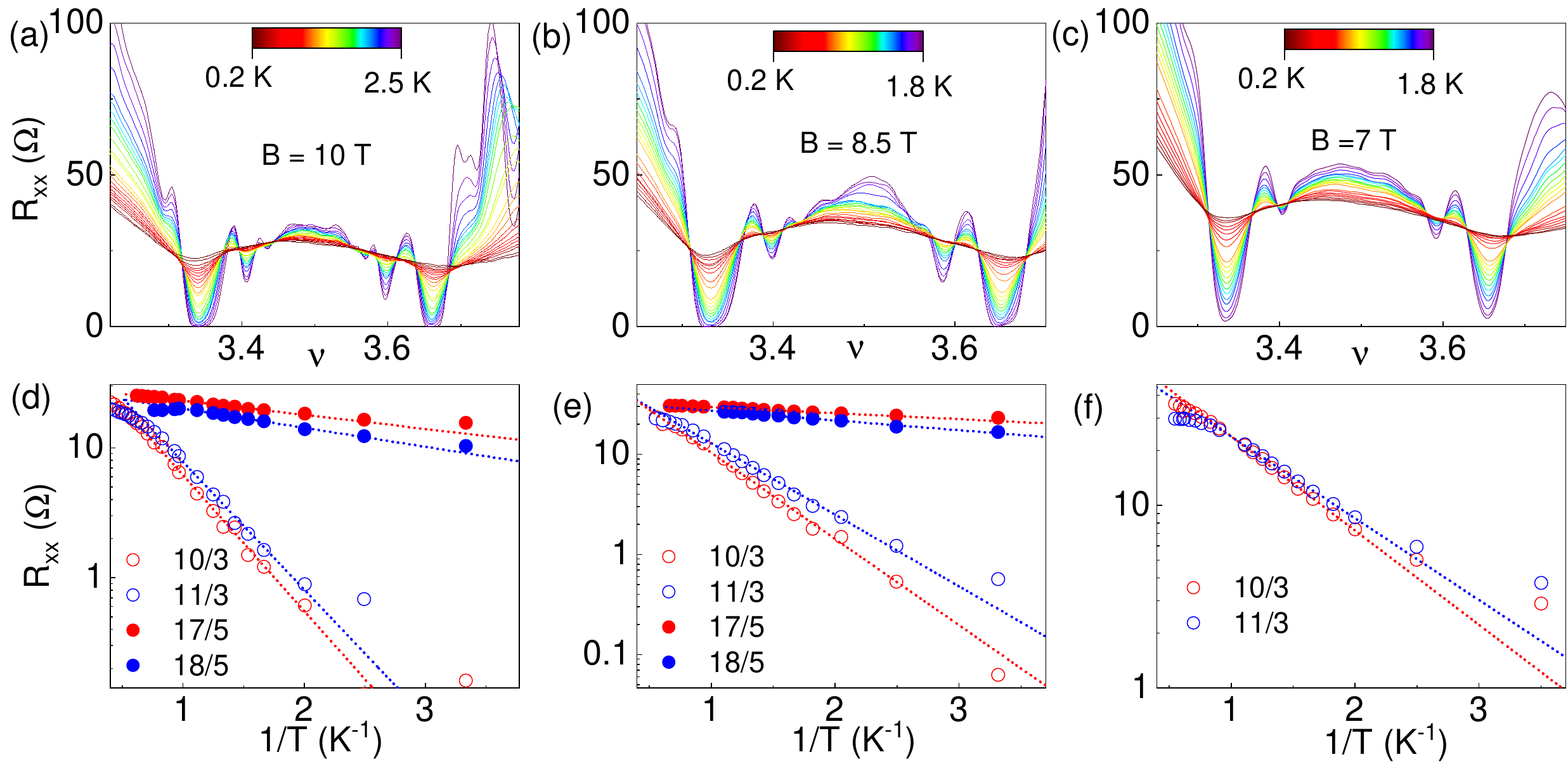}
	\caption{\textbf{Activation gaps.} Plots of $R_{xx}$ versus $\nu$ at a few representative temperatures measured at (a) $B=10$~T, (b) $B=8.5$~T, and (c) $B = 7$~T. Arrhenius fits to $R_{xx}$ at (d) $B=10$~T, (e) $B=8.5$~T, and (f) $B = 7$~T. The data are for $|D|=0.1$~V/nm.}
	\label{fig:figS2}
\end{suppfigure}

\section{Controlled violation of PH symmetry of FQHs.}
Fig.~\ref{fig:figS3} shows the temperature dependence of $R_{xx}$ for various $D$ in the vicinity of Landau level mixing for filling factor $\nu=2$ and $3$. The blue and green shaded regions highlight the resistance dips for $\nu=7/3$ and $8/3$ FQHs. These states exhibit PHS for $|D|=0.77$~V/nm and $0.87$~V/nm, which is far from the crossing region. However, in the range $0.82<|D|<0.86$~V/nm, the $R_{xx}$ minima of the hole state $\nu=8/3$ gradually diminishes. For this range of $|D|$, $LL_B^{2+} \uparrow$ and $LL_M^{0+}\uparrow$ come close and cross each other, resulting in the observed violation of PH symmetry. Activation gaps for these FQHs are shown in Fig.~4(c) of the main manuscript.

\begin{suppfigure}
	\includegraphics[width=0.7\columnwidth]{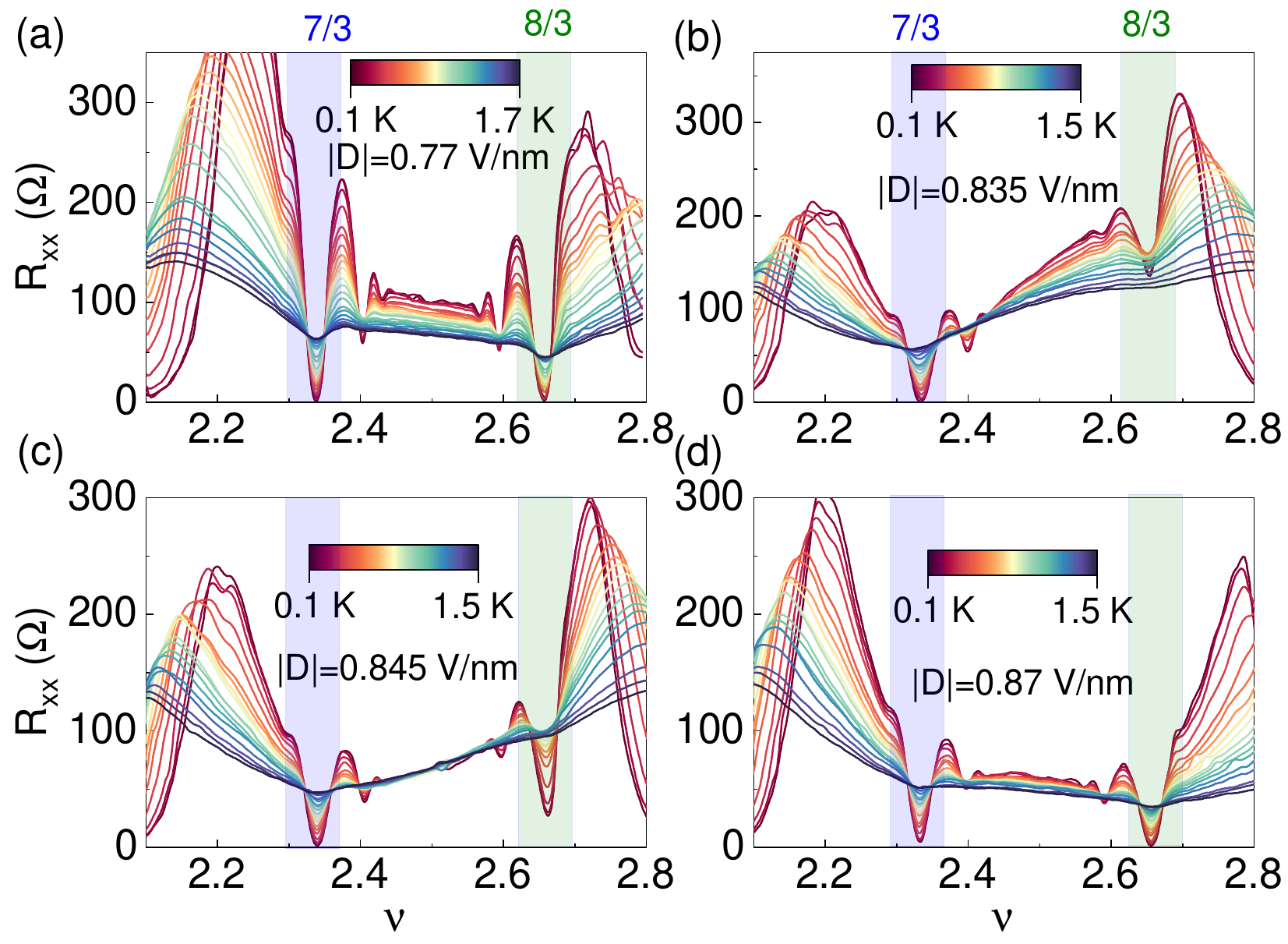}
	\caption{\textbf{ Controlled breaking of PH symmetry of FQHs in the vicinity of Landau level crossing:} Plots of $R_{xx}$ versus $\nu$ at different temperatures for (a) $|D|=0.77$~V/nm (b) $|D|=0.835$~V/nm (c) $|D|=0.845$~V/nm (d) $|D|=0.87$~V/nm. The data is taken for $B=10$~T. The violation of the PH symmetry between $\nu=7/3$ (marked by blue shaded rectangle) and $\nu=8/3$ (marked by green shaded rectangle) is apparent from the plots.}
	\label{fig:figS3}
\end{suppfigure}

\section{Simulated Landau level spectrum of ABA trilayer graphene as a function of interlayer potential $\Delta_1$ and LL energy $E$.}

Fig. \ref{fig:figS4} shows the band structure of ABA trilayer graphene calculated at (a) $\Delta_1=0$~meV and (b) $\Delta_2=35$~meV. The bilayer-like band (BLL) is highlighted in green, while the monolayer-like band (MLL) is in brown. At $\Delta_1=0$~meV, mirror symmetry protects the two bands from mixing. At a finite $\Delta_1$, mirror symmetry breaks, and two bands hybridize. Consequently, the MLL band splits into three distinct sections ($M-$,$M+$,$M2$). With increasing $\Delta_1$, $M-$, $M2$ bands move higher in energy.

Fig. \ref{fig:figS4}(c) shows the calculated landau level spectrum of ABA trilayer graphene as a function of $\Delta_1$ and $E$.
Multiple LL crossings between the MLL and BLL landau levels are evident. The shaded rectangles mark the region where the PH symmetry violation is observed and explored.

\begin{suppfigure}[h]
	\includegraphics[width=0.6\columnwidth]{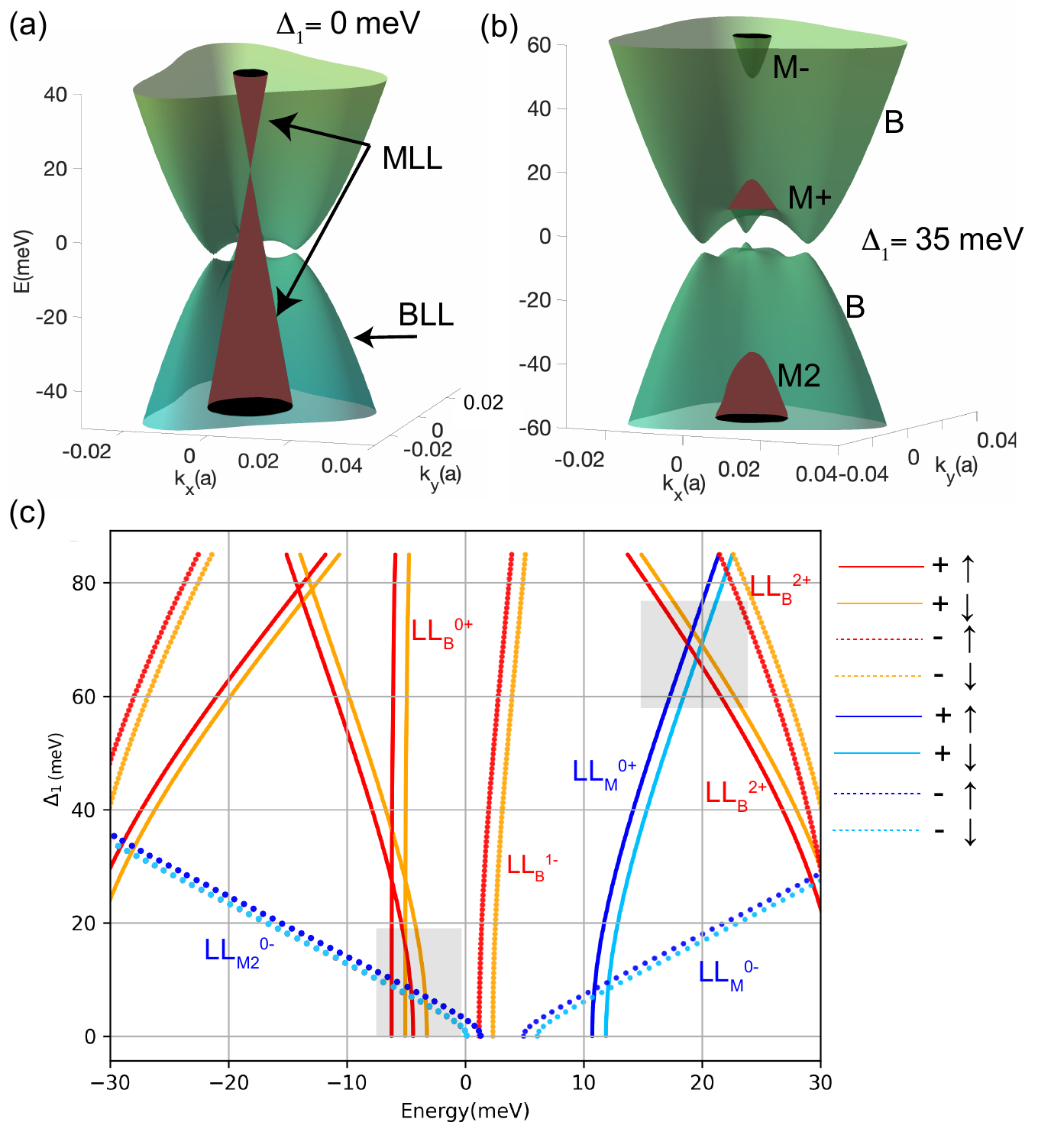}
	\caption{\textbf{Band structure and LL evolution of ABA trilayer graphene as a function of $\Delta_1$.} Calculated band structure of ABA trilayer graphene at (a) $\Delta_1=0$~meV (b) $\Delta_1=35$~meV. Here, the inside band (brown-shaded region) is the monolayer-like band (MLL), and the outside band (green-shaded region) is the bilayer-like band (BLL). (c) Simulated Landau level plot of ABA trilayer graphene as a function of $\Delta_1$ and $E$. The plot is simulated for $B=10$~T. Here, red (up-spin) and orange color(down-spin) LLs originate from the BLL band. Cyan (down spin) and blue color (up spin) LLs originate from the MLL band. The shaded gray rectangle marks the region where the observed PH symmetry violation of FQHs is explored in this study.}
	\label{fig:figS4}
\end{suppfigure}

\section{Evolution of PH asymmetry of FQHs with $B$.}

Fig. \ref{fig:figS5} shows the contour plot of $G_{xx}$ as a function of $D$ and filling factor $\nu$ measured at different magnetic fields $B$ specified in the plot. At $B=12$~T, the fractional quantum Hall states exhibit PH symmetry about half-filling across the entire $\nu-D$ plane. However, at $B=11$~T, certain regions deviate from this symmetry, which are marked using distinct symbols. As the magnetic field is reduced to $B=10$~T, these PH symmetry-breaking regions shift towards lower values of $D$. At an even lower field $B=8$~T, the asymmetry further extends to significantly lower $D$ values.

To understand this behavior, we simulated the Landau level plot as a function of $\Delta_1$ and $E$ at the different $B$. The plots are shown in Fig. \ref{fig:figS5}(e-h) respectively. We see a similar trend in the Landau level crossings between $LL_M^{0+}$ of the MLL band and $LL_B^{2+}$ of the BLL band. As the magnetic field decreases, these crossings shift towards lower $\Delta_1$.

Fig. \ref{fig:figS6} is a quantitative comparison of the $D_{experiment}$ and $D_{theory}$ as a function of the magnetic field $B$. Here, $D_{experiment}$ marks the value around which PH symmetry violation of FQHs is observed between filling factor $\nu=2$ and $3$. $D_{theory}$ is value at the crossing of $LL_M^{0+}\uparrow$ and $LL_B^{2+}\uparrow$ LLs calculated from the simulated plot Fig. \ref{fig:figS5}(e-h). The strong agreement between the experimental and theoretical values of $D$ confirms that the observed violation of PH symmetry arises due to the Landau level mixing of two bands.
\begin{suppfigure}[h]

	\includegraphics[width=0.9\columnwidth]{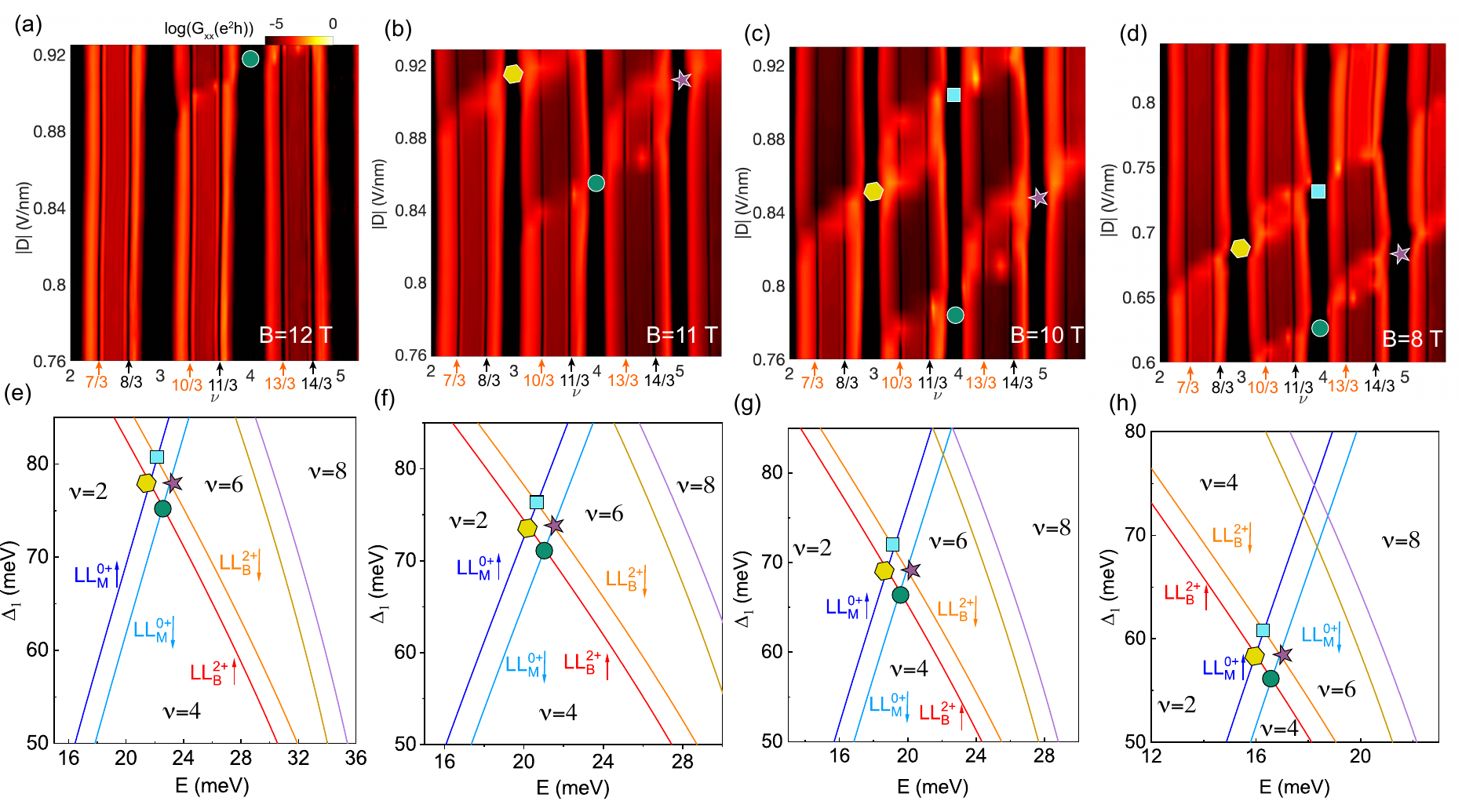}
	\caption{\textbf{ Evolution of PH symmetry violation with perpendicular magnetic field $B$:} Contour plot of $G_{xx}$ as a function of $D$ and filling factor $\nu$ measured at (a) $B=12$~T, (b) $B=11$~T, (c) $B=10$~T, (d) $B=8$~T. Landau level plots as a function of $\Delta_1$ and $E$ calculated at (e) $B=12$~T, (f) $B=11$~T, (g) $B=10$~T, (h) $B=8$~T. Each symbol marks a crossing point of Landau levels.}
	\label{fig:figS5}
\end{suppfigure}

\begin{suppfigure}[h]
	\includegraphics[width=0.4\columnwidth]{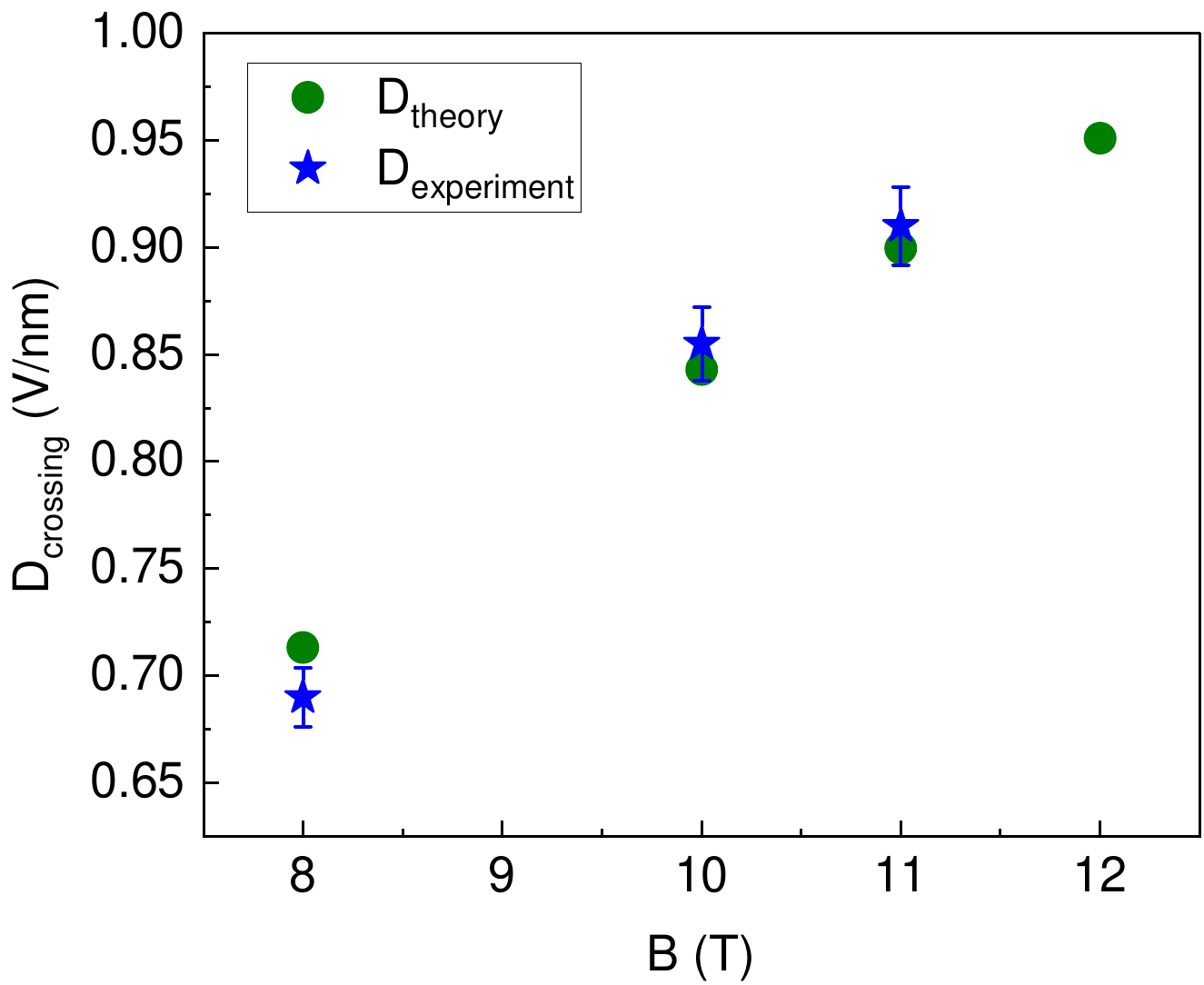}
	\caption{\textbf{ Comparison of $D_{theory}$ and $D_{experiment}$:} Plot of measured $D_{experiment}$ around which PH symmetry violation is experimentally observed and $D_{theory}$ estimated from simulated LLs as a function of magnetic field $B$.}
	\label{fig:figS6}
\end{suppfigure}

\section{Visibility of LL crossings at high temperatures}

Fig. \ref{fig:figS7} shows the comparison contour plot of $G_{xx}$ as a function of $\nu$ and $|D|$ taken at (a) $T=0.5$~K and (b) $T=2$~K. The Landau level crossings are better observed at $T=2$~K. Each LL (measured at $T=2$~K Fig. \ref{fig:figS7}(b)) and observed PH symmetry violation (measured at $T=0.5$~K Fig. \ref{fig:figS7}(a)) is marked by the same symbol for consistency. This further confirms the direct correlation between the $D$ range where PH asymmetry and Landau level crossings between MLL and BLL bands are seen.

\begin{suppfigure}[h]
	\includegraphics[width=0.7\columnwidth]{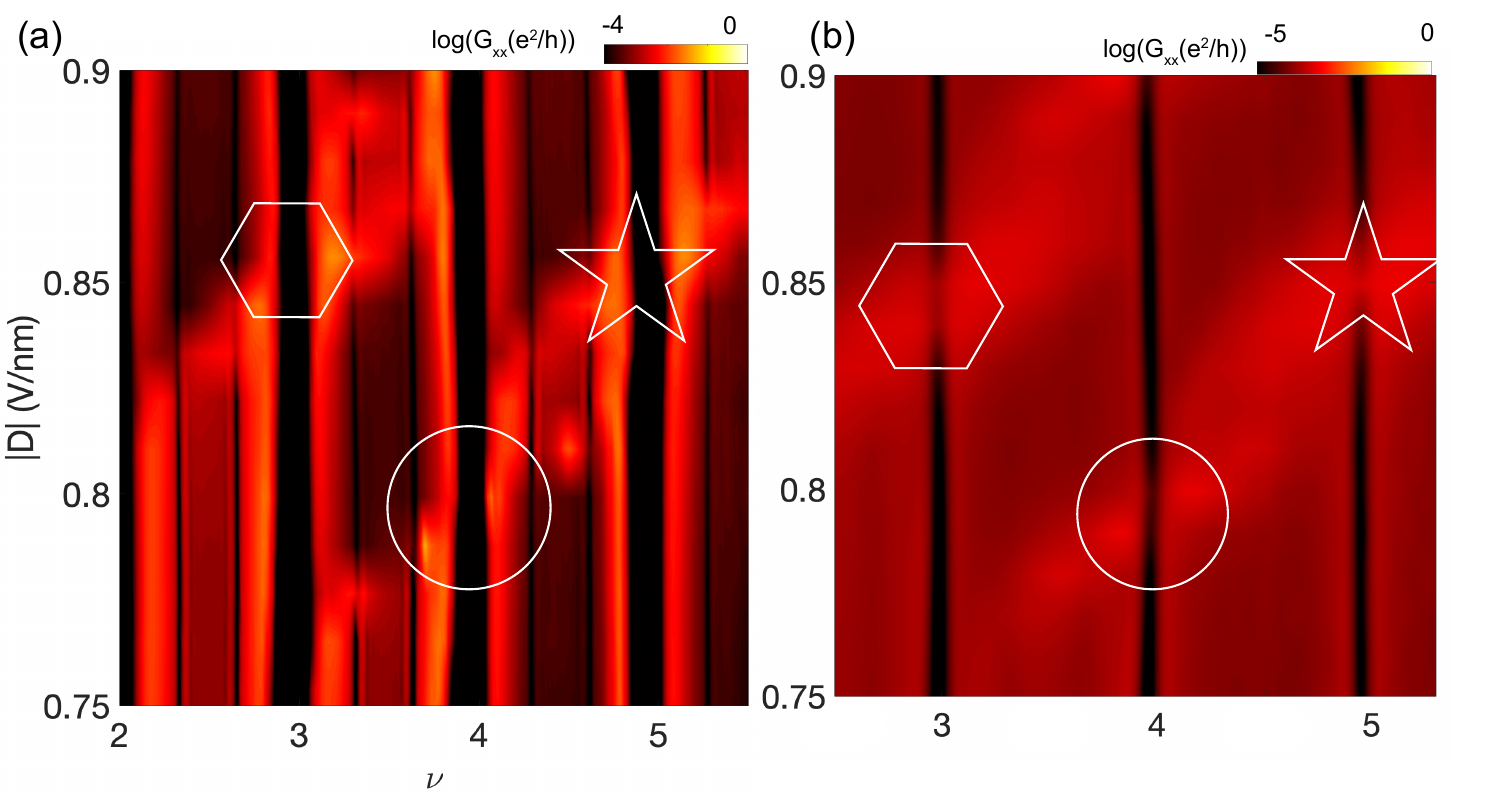}
	\caption{\textbf{Visibility of Landau level crossings at $T=2$~K}. Contour plot of $G_{xx}$ as a function of $|D|$ and filling factor $\nu$ measured at (a) $T=0.5$~K, (b) $T=2$~K. Each symbol marks the LL crossing and observed PH symmetry violation of FQHs.}
	\label{fig:figS7}
\end{suppfigure}

\section{Line scans of $R_{xx}$ in the vicinity of Landau level mixing.}

Fig. \ref{fig:figS8} shows the line scans of $R_{xx}$ as a function of $\nu$ measured at different values of $|D|$. Here, blue (pink) dashed lines mark the dips in the $R_{xx}$ for particle (hole) FQHs at (a) $\nu=-7/3$ and $-8/3$, (b) $\nu=7/3$ and $8/3$, (c) $\nu=10/3$ and $11/3$, (d) $\nu=13/3$ and $14/3$. Table \ref{table:tablephviola} shows the comprehensive summary of PH symmetry violations.

\begin{suppfigure}[h]
	\includegraphics[width=0.7\columnwidth]{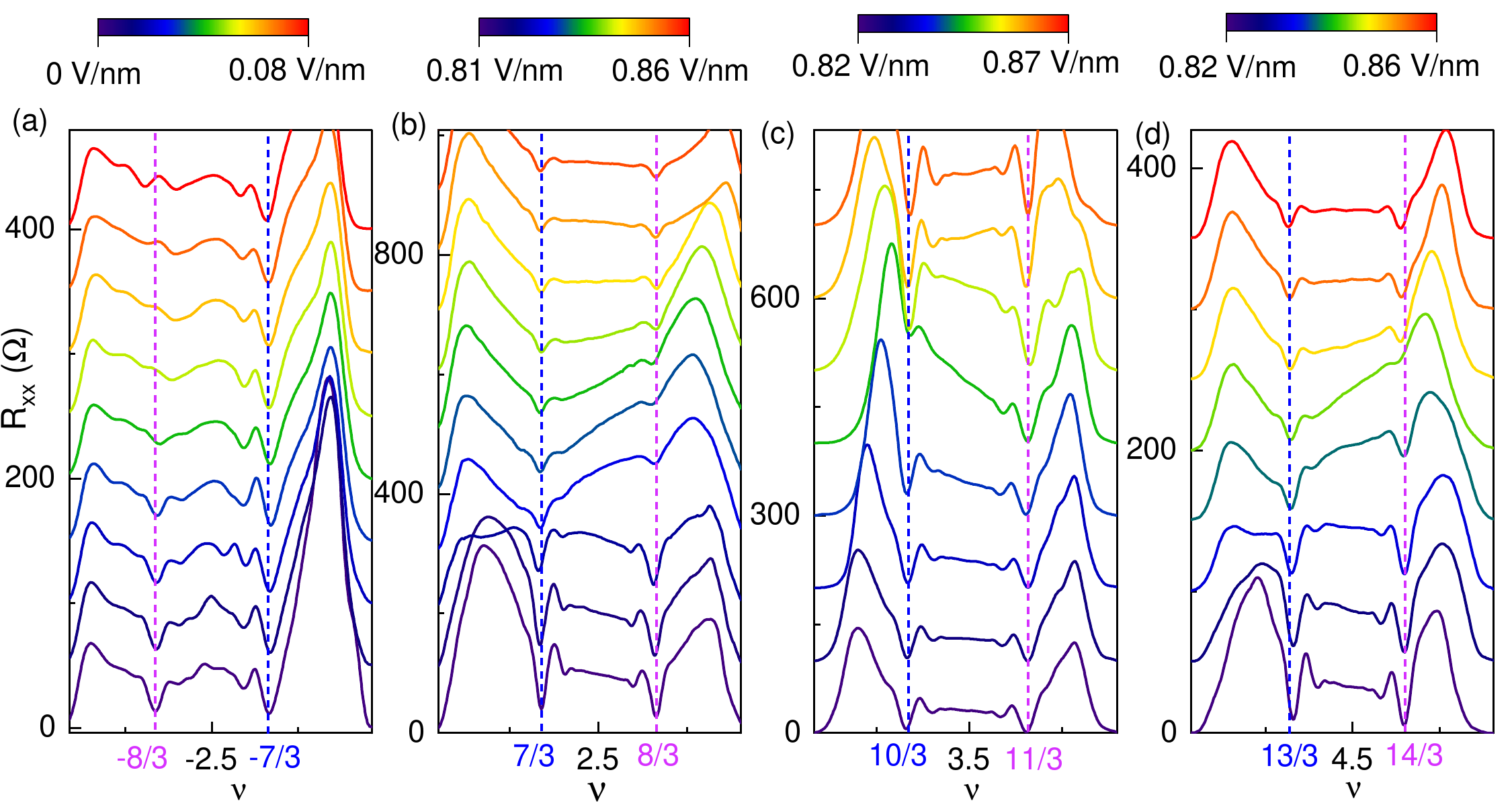}
	\caption{\textbf{Line scans of $R_{xx}$ in the vicinity of Landau level crossing.} Plot of $R_{xx}$ as a function of $\nu$ for various values of $|D|$ in the vicinity of landau level crossing between  (a) $LL_{M2}^{0-}\uparrow$ and $LL_{B}^{1+}\uparrow$, (b) $LL_{M}^{0+}\uparrow$ and $LL_{B}^{2+}\uparrow$, (c) $LL_{M}^{0+}\uparrow$ and $LL_{B}^{2+}\uparrow$, (d) $LL_{M}^{0+}\downarrow$ and $LL_{B}^{2+}\downarrow$. Here, blue dashed lines mark the particle FQHs, and pink dashed line marks the hole conjugate FQHs. $R_{xx}$ plots at different $|D|$ are offset by constant value in each plot for representation.}
	\label{fig:figS8}
\end{suppfigure}

\begin{table}[h]
	\centering
	\resizebox{\columnwidth}{!}{
		\begin{tabular}{| p{.25\textwidth} | p{.25\textwidth} | p{.25\textwidth} |p{.25\textwidth} |}
			\hline
			$\nu$ & Observed FQHs &  Nature of Landau level crossing &  FQHs most affected \\
			\hline
			$-2$ and $-3$  & $-7/3$ and $-8/3$ & $LL_{M2}^{0-}\uparrow$ and $LL_{B}^{1+}\uparrow$ &  $-8/3$ \\
			\hline

			$-3$ and $-4$  & $-10/3$ and $-11/3$ & $LL_{M2}^{0-}\uparrow$ and $LL_{B}^{1+}\downarrow$ &  $-11/3$ \\
			\hline

			$-4$ and $-5$  & $-13/3$ and $-14/3$ & $LL_{M2}^{0-}\downarrow$ and $LL_{B}^{1+}\uparrow$ &  $-11/3$ \\
			\hline

			$2$ and $3$  & $7/3$ and $8/3$ & $LL_{M}^{0+}\uparrow$ and $LL_{B}^{2+}\uparrow$ &  $8/3$ \\
			\hline

			$3$ and $4$  & $10/3$ and $11/3$ & $LL_{M}^{0+}\uparrow$ and $LL_{B}^{2+}\downarrow$ &  $10/3$ \\
			\hline

			$3$ and $4$  & $10/3$ and $11/3$ & $LL_{M}^{0+}\downarrow$ and $LL_{B}^{2+}\uparrow$ &  $10/3$ \\
			\hline

			$4$ and $5$  & $13/3$ and $14/3$ & $LL_{M}^{0+}\downarrow$ and $LL_{B}^{2+}\downarrow$ &  $14/3$ \\
			\hline

	\end{tabular}}
	\caption{Comprehensive summary of PH symmetry violation in different filling factors $\nu$ due to mixing between BLL and MLL Landau levels. }
	\label{table:tablephviola}
\end{table}

\section{Quantitative analysis of controlled violation of PH symmetry of FQHs in $\nu=3$ and $4$.}

Fig. \ref{fig:figS9}(a) shows the contour plot of $G_{xx}$ as a function of $|D|$ and $\nu$ between filling factor $\nu=3$ and $4$. The measured activation gaps as a function of $|D|$ for FQH states $\nu=10/3$ and $11/3$ are shown in Fig. \ref{fig:figS9}~(b). The shaded regions in Fig. \ref{fig:figS9}(b) mark the region of observed PH symmetry violation (regions where the measured FQH activation gaps of $\nu=10/3$ drops to zero.

\begin{suppfigure}[h]
	\includegraphics[width=0.7\columnwidth]{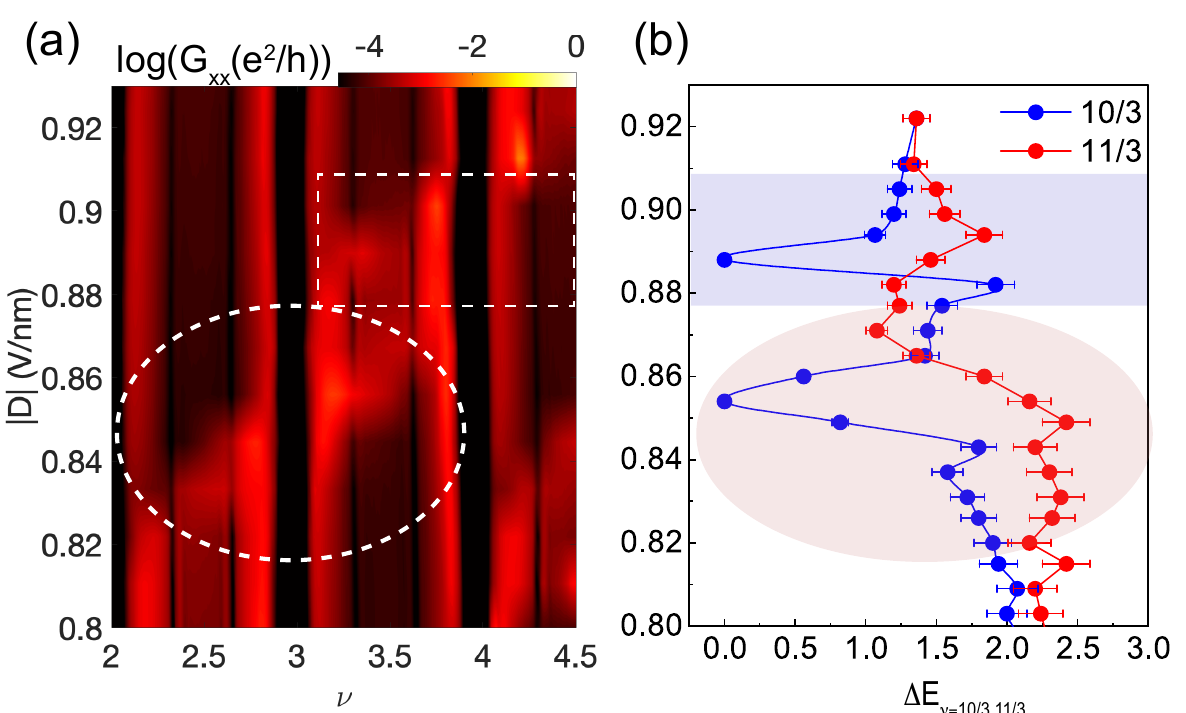}
	\caption{\textbf{ Quantitative analysis of PH symmetry violation of FQHs in $\nu=3$ and $4$}. Contour plot of $G_{xx}$ as a function of $|D|$ and $\nu$ between (a) $\nu=3$ and $4$. Plot of measured activation gaps $\Delta$ as a function of $|D|$ for (b) $\nu=10/3$ and $11/3$. Here, dashed rectangle and circle marks the region where PH asymmetry is observed.}
	\label{fig:figS9}
\end{suppfigure}

\section{Displacement field dependent Landau level Fan diagram of aba trilayer graphene.}

Fig. \ref{fig:figSS10}(a-c) shows the contour plot of $G_{xx}$ as a function of $B$ and number density $n$ at displacement fields (a) $|D|=0$~V/nm, (b) $|D|=0.3$~V/nm, (c) $|D|=0.5$~V/nm. Here, $LL_{M}^0$ are the Landau levels originating from the MLL band, and $LL_{B}^{2,3,4}$ are the Landau levels originating from the BLL bands. At $D=0$~V/nm (Fig. \ref{fig:figSS10}(a)), we observe four MLL LLs crossing multiple BLL like LLs at $n\approx0.5\times10^{19}$~m$^{-2}$, consistent with the calculated bandstructure of ABA trilayer graphene where charge neutrality point of MLL band is located at $E\approx18$~meV.
For $D>0$~V/nm (Fig \ref{fig:figSS10}(b),(c)), $LL_{M}^{0-}$ shifts to higher energies while $LL_{M}^{0+}$ remains close to its original position. This behavior aligns well with bandstructure for $\Delta_1>0$~meV (Fig \ref{fig:figS4}(b)), where conduction band of MLL band (M-) shifts to higher energies while valence band(M+) remains unaffected.

Additionally, we have also added the simulated Landau level spectrum as a function of $n$ and $B$ at different $\Delta_1$ for comparison(Fig \ref{fig:figSS10}(d-f)). There is strong agreement with measured Landau level fandiagrams (Fig \ref{fig:figSS10}(a-c)).

\begin{suppfigure}[h]
	\includegraphics[width=\columnwidth]{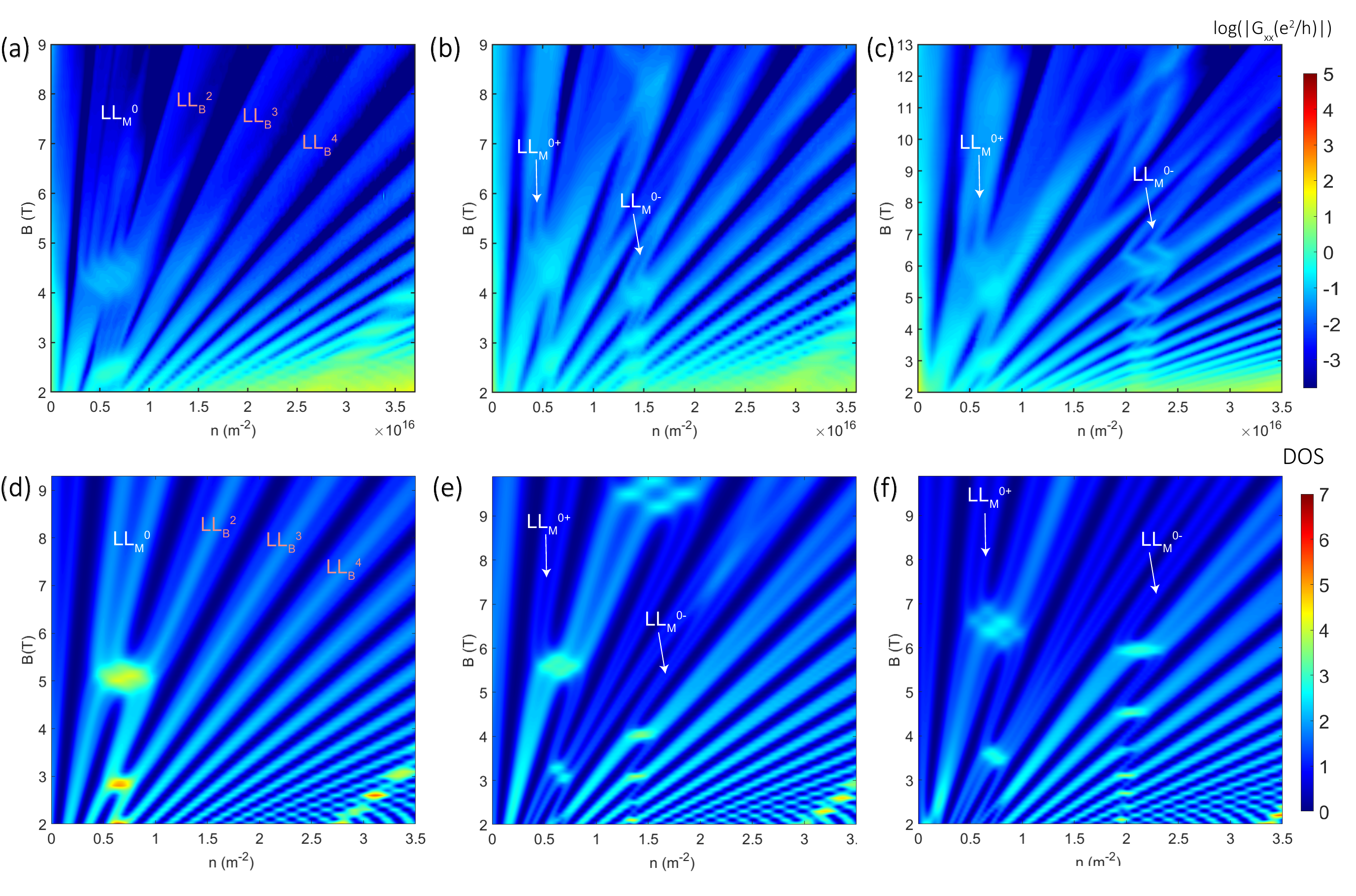}
	\caption{\textbf{$D$ dependent fan diagram.} Contour plot of $G_{xx}$ as a function of $B$ and $n$ measured at (a) $|D|=0$~V/nm, (b) $|D|=0.3$~V/nm, (c) $|D|=0.5$~V/nm. Calculated Landau level spectrum of the ABA TLG at (d) $\Delta_1=0$ meV, (e) $\Delta_1=25$~meV, (f) $\Delta_1=40$~meV. }
	\label{fig:figSS10}
\end{suppfigure}

\section{Particle-Hole Symmetry in device~3}

Fig. \ref{fig:figS10} shows the plot of $G_{xx}$ and $G_{xy}$ versus $\nu$ for device3. One can see clear particle-hole symmetry around half filling between $\nu = 2$ and $3$ and between $\nu = 3$ and $4$.


Fig. \ref{fig:figS11}(a) shows the contour plot of $R_{xx}$ as a function of filling factor $\nu$ measured at $B=12$ T. The arrows indicate the FQH states. Vanishing of the resistance minima at similar temperatures for particle and hole states around half-filling confirms the existence of particle-hole symmetry in device3 at low displacement fields. To quantify it further, we have measured activation gaps as a function of the magnetic field shown in Fig. \ref{fig:figS11}(b). The measured gaps exhibit significant reflection symmetry around $B_{eff} = 0$ T, further confirming the particle-hole symmetry around the half-filling.

The gaps for $\nu = 2+2/5$ and $2+3/5$ have a $\sqrt{B}$-dependence (Fig. \ref{fig:figS11}(b)). As discussed in the main manuscript, we fit the data with the equation for spinless CF,  $\Delta = \hbar eB_{eff}/m_{eff} - \Gamma$~\cite{PhysRevLett.92.156401, PhysRevLett.122.137701}. Here, $\Gamma$ is the disorder-induced Landau level broadening and $m_{eff}=\alpha m_e\sqrt{(2p+1)B_{eff}}$ is the effective CF mass  ($\alpha$ is the effective mass parameter and $m_e$ the mass of free electrons). We get $\alpha_{2+2/5} = 0.10 \pm 0.0032$, $\alpha_{2+3/5} = 0.10\pm 0.0029$, $\Gamma_{2+2/5} =7.79\pm 0.45$~K,  and  $\Gamma_{2+3/5} =7.64\pm 0.55$~K.

The gaps for $\nu = 2+1/3$ and  $2+2/3$ can be fitted equally well with either a linear $B$-dependence or a $\sqrt{B}$-dependence.
We quantify these gaps using  $\Delta_S=\frac{1}{2}\mu_B\mathbf{g}(2p+1)B_{eff}-\Gamma$ with $\mathbf{g}$ the effective Land\'e g-factor. Fits to the data yields $\mathbf{g}_{2+1/3} = 3.81\pm 0.31$,  $\mathbf{g}_{2+2/3} = 3.93\pm 0.37$, $\Gamma_{2+1/3} = 7.47\pm 1.07$~K  and $\Gamma_{2+2/3} = 7.70\pm 2.04$~K. On the other hand, the disorder broadening extracted from the $\sqrt B$ fits is $\Gamma \sim 22$~K for both particle and hole conjugate states.

\begin{suppfigure}[h]
	\includegraphics[width=0.8\columnwidth]{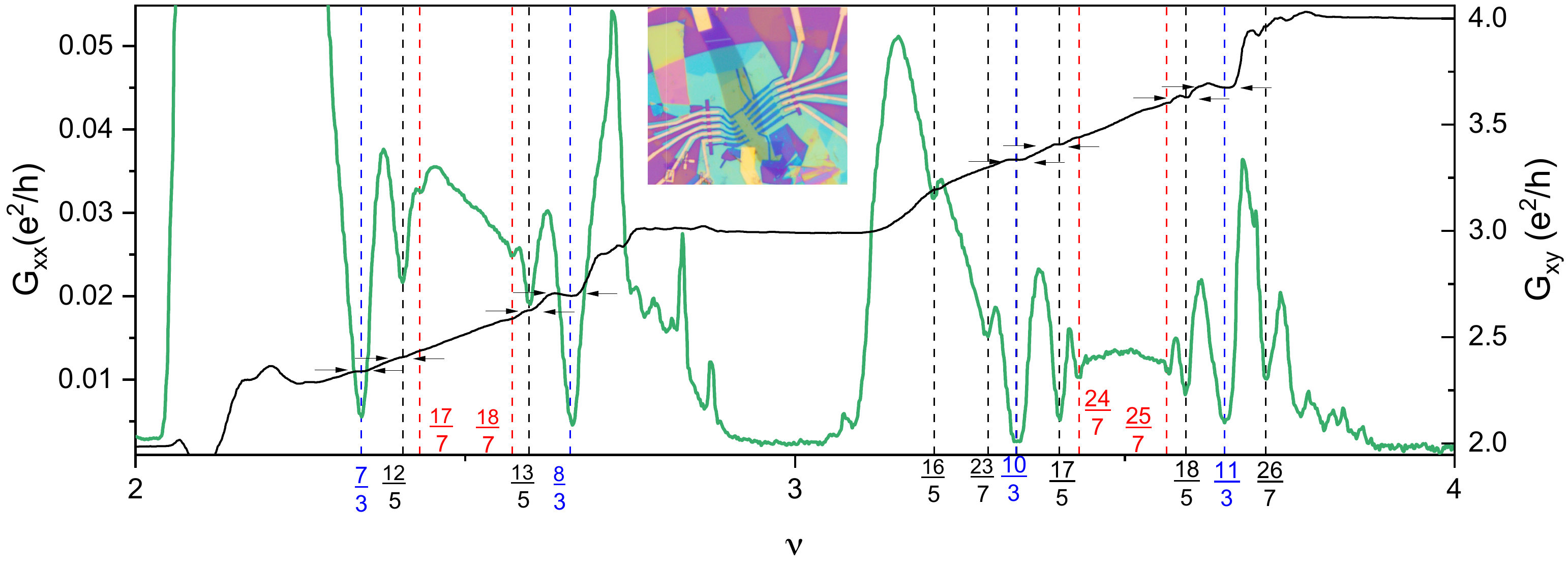}
	\caption{\textbf{PH symmetry in device3.} Plot of $G_{xx}$ versus $\nu$ showing the particle-hole symmetry of the FQH states at $T=20$ mK and $|D|\approx0$ for the device2. Inset shows the optical image of the device3.}
	\label{fig:figS10}
\end{suppfigure}
\begin{suppfigure}[h]
	\includegraphics[width=0.8\columnwidth]{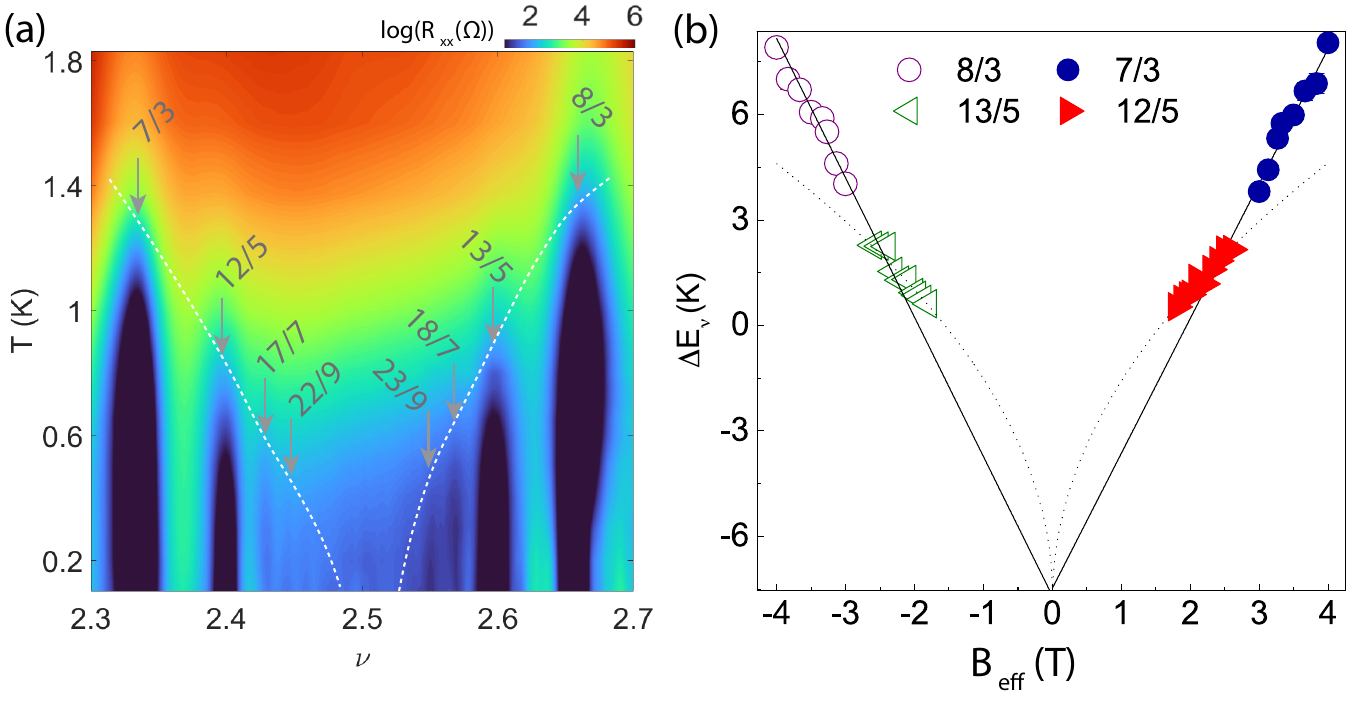}
	\caption{\textbf{Particle hole symmetry and activation gaps of FQHs in device3.} (a) Contour map of $R_{xx}$ as a function of $T$ and $\nu$. The dotted white line outlines the $T$-range over which the FQH states persist. The data were taken at $B =12$~T. (b)  Excitation gaps as a function of $B_{eff}$ for FQH states. The solid black lines are linear fits to the data for $\nu = 8/3$ and $7/3$. The dotted gray lines are $\sqrt{B}$-fits of the data for $\nu = 13/5$ and $12/5$.}
	\label{fig:figS11}
\end{suppfigure}

\section{Comparison of \texorpdfstring{$B$}{B} and \texorpdfstring{$\sqrt{B}$}{sqrt B} dependence of \texorpdfstring{$\Delta$}{Delta}}

Fig. \ref{fig:figS12} shows the fitting of activation gaps for $\nu = 10/3$ and $11/3$ and $\nu = 7/3$ and $8/3$ using  $\sqrt B$-
(black dashed line) and linear-$B$ dependence (red dashed lines). The disorder broadening extracted from the $\sqrt B$ fits $\Gamma \sim 14$~K ($\nu = 7/3$ and $8/3$) and $\Gamma \sim 12$~K ($\nu = 10/3$ and $11/3$) which is similar for both particle and hole conjugate states. The parameters extracted from $\sqrt{B}$ to the data points are given in Table 1 and 2 of the main text. The plots are shown for device~1.
\begin{suppfigure}[h]
	\includegraphics[width=0.8\columnwidth]{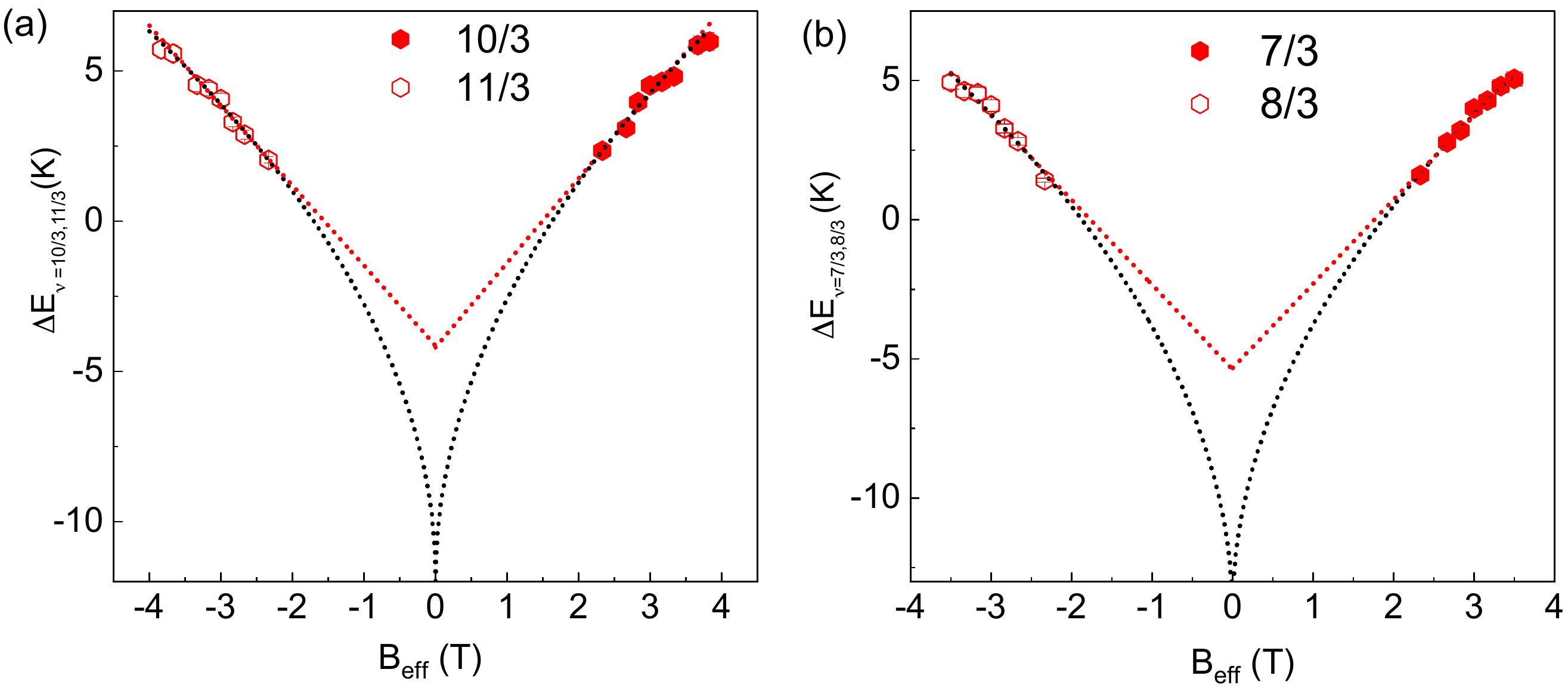}
	\caption{\textbf{ Comparison of $\sqrt B$ and linear-$B$ behavior for $\nu =n+1/3$ and $n+2/3$:} Plots of activation energy $\Delta$ as a function of $B_{eff} = B/(2p+1)$ for (a) $\nu$ = 10/3 and 11/3 (b) $\nu$ = 7/3 and 8/3. The black dashed lines are fits of the data points using the $\sqrt B$ equation for activation gaps. Red dashed lines are fit using the Zeeman term, which is linear in $B$.}
	\label{fig:figS12}
\end{suppfigure}

\section{Distribution of wavefunction at the atomic sites in various LLs of ABA TLG.}

Fig 3(e) of the main manuscript shows the detailed schematics illustrating the localization of the wavefunction of LLs at the different atomic sites in ABA TLG. The schematics are derived from the wavefunctions detailed in Table \ref{table:tableS1}. These wavefunctions on different atomic sites are taken from the recent publication \cite{shimazaki2016landau,PhysRevB.105.115126}. Notably, the schematics and energy versus $\Delta_1$ plot (Fig \ref{fig:figS4} (c)) of ABA TLG reveal that FQHs hosted by $LL_{M}^{0+}$ are single component. The only significant mixing into $LL_{M}^{0+}$ originates from the  $LL_B^{2+}$ LL of bilayer like band.

\begin{table}[h]
	\centering
	\resizebox{\columnwidth}{!}{
		\begin{tabular}{| p{.50\textwidth} | p{.50\textwidth} |}
			\hline
			$LL$   & Wavefunction \\
			\hline

			$LL_{M}^{0+}$ & $\ket{0} \circledast \frac{\ket{A_1}-\ket{A_3}}{\sqrt{2}} \circledast \ket{K}$  \\
			\hline

			$LL_{M}^{0-}$ & $\ket{0} \circledast \frac{\ket{B_1}-\ket{B_3}}{\sqrt{2}} \circledast \ket{K^{'}}$  \\
			\hline

			$LL_{B}^{2+}$ & $a_2^{b+}\ket{0} \circledast \frac{\ket{A_1}+\ket{A_3}}{\sqrt{2}}+b_2^{b+} \ket{2} \circledast \ket{B_2} \circledast \ket{K}$  \\
			\hline

	\end{tabular}}
	\caption{The correspondence between the LL and the wavefunction at $D\approx0$ V/nm in ABA TLG.}
	\label{table:tableS1}
\end{table}

\section{Effective mass, Land\'e g-factor and disorder broadening in pristine TLG.}

\ref{table:table1} and \ref{table:table2} shows the extracted CF effective mass parameter, Land\'e g-factor, and the disorder broadening from the fits shown in the Fig 2(c) and (d) of the main manuscript between filling factor $\nu=2$ and $3$, $\nu=3$ and $4$. The obtained values shows excellent PH symmetry of FQHs.
\begin{table}[hb]
	\centering
	\resizebox{\columnwidth}{!}{
		\begin{tabular}{| p{.20\textwidth} | p{.20\textwidth} | p{.2\textwidth} | p{.20\textwidth} |p{.20\textwidth} |}
			\hline
			$\nu$  &  $7/3$ &  $8/3$ & $12/5$ & $13/5$ \\
			\hline

			$\alpha$ (from $\sqrt{B}$ fit) & $0.078 \pm 0.003$ & $0.077 \pm 0.006$ &  $0.238 \pm 0.048$ & $0.270 \pm 0.15$ \\
			\hline

			$g_{eff}$ (from $B$ fit)  & $3.00 \pm 0.16$ & $3.02 \pm 0.30$& -- & -- \\ \hline

			$\Gamma~(K)$ (from $\sqrt{B}$ fit) & $14.1 \pm 0.78$ K & $14.2 \pm 1.1$ & $3.33 \pm 0.76$ K  & $2.85\pm 1.8$ K \\ \hline

			$\Gamma~(K)$ (from $B$ fit) & $5.33 \pm 0.49$ K & $5.38 \pm 0.89$ & -- & --\\ \hline

	\end{tabular}}
	\caption{Values of effective mass, effective $g$ and disorder broadening parameters for FQHs between LL $\nu =2$ and $3$.}
	\label{table:table1}
\end{table}

\begin{table}[hb]
	\centering
	\resizebox{\columnwidth}{!}{
		\begin{tabular}{| p{.20\textwidth} | p{.20\textwidth} | p{.2\textwidth} | p{.20\textwidth} |p{.20\textwidth} |}
			\hline
			$\nu $   &  $10/3$ &  $11/3$ & $17/5$ & $18/5$ \\
			\hline

			$\alpha$ (from $\sqrt{B}$ fit) & $0.085 \pm 0.004$ & $0.088 \pm 0.003$ &  $0.1778 \pm 0.018$ & $0.1794 \pm 0.017$ \\
			\hline

			$g_{eff}$ (from $B$ fit)  & $2.79 \pm 0.14$ & $2.65 \pm 0.12$ & -- & -- \\ \hline

			$\Gamma~(K)$ (from $\sqrt{B}$ fit) & $12 \pm 0.64$ K & $11.9 \pm 0.62$ & $4.3 \pm 0.51$ K  & $4.1 \pm 0.47$ K \\ \hline

			$\Gamma~(K)$ (from $B$ fit) & $4.2 \pm 0.36$ K & $4.15 \pm 0.37$ & --  & -- \\ \hline
	\end{tabular}}
	\caption{Values of effective mass, effective g, and disorder broadening parameters for FQHs between LL $\nu = 3$ and $4$.}
	\label{table:table2}
\end{table}

\clearpage

\bibliography{PHrefrence}

\end{document}